\documentclass[draftcls,onecolumn,11pt]{IEEEtran}
\usepackage[T1]{fontenc}% optional T1 font encoding
\usepackage{rotating,graphics,psfrag,epsfig}

\usepackage{cite}
\usepackage{flushend}
\usepackage{xcolor}
\usepackage{subfigure}
\usepackage{amsmath}
\usepackage{amsfonts}
\usepackage{amssymb}
\usepackage{algorithmic}
\usepackage{algorithm}
\usepackage{array}

\usepackage{acronym}
\usepackage{url}       % Written by Donald Arseneau
\acrodef{DFT}{discrete Fourier transform}
\acrodef{IFFT}{inverse fast Fourier transform}
\acrodef{FFT}{fast Fourier transform}
\acrodef{SNR}{signal-to-noise ratio}
\acrodef{CP}{cyclic prefix}
\acrodef{ISI}{inter symbol interference}
\acrodef{mm-wave}{millimeter-wave}
\acrodef{cm-wave}{centimeter-wave}
\acrodef{MIMO}{multiple-input-multiple-output}
\acrodef{SNR}{signal-to-noise ratio}
\acrodef{FIM}{Fisher information matrix}
\acrodef{CRB}{Cram\'{e}r-Rao bound}
\acrodef{UWB}{ultra-wide bandwidth}
\acrodef{GPS}{global positioning system}
\acrodef{BF}{beamforming}
\acrodef{QoS}{quality-of-service}
\acrodef{LOS}{line-of-sight}
\acrodef{NLOS}{non-line-of-sight}
\acrodef{OLOS}{obstructed-line-of-sight}
\acrodef{RF}{radio-frequency}
\acrodef{EXIP}{extended invariance principle}
\acrodef{DLS}{damped least-squares}
\acrodef{CDF}{cumulative distribution function}
\acrodef{MPCs}{multi-path components}
\acrodef{ML}{maximum likelihood}
\acrodef{MS}{mobile station}
\acrodef{WLS}{weighted least squares}
\acrodef{LMA}{Levenberg-Marquardt algorithm}
\acrodef{GNA}{Gauss-Newton algorithm}
\acrodef{BS}{base station}
\acrodef{ADC}{analog-to-digital-converter}
\acrodef{AOA}{angle-of-arrival}
\acrodef{DOA}{direction-of-arrival}
\acrodef{AOD}{angle-of-departure}
\acrodef{TOA}{time-of-arrival}
\acrodef{TDOA}{time-difference-of-arrival}
\acrodef{ULA}{uniform linear array}
\acrodef{PSD}{positive semidefinite}
\acrodef{EFIM}{equivalent Fisher information matrix}
\acrodef{FB}{fractional bandwidth}
\acrodef{REB}{rotation error bound}
\acrodef{PEB}{position error bound}
\acrodef{SDL}{sensor delay line}
\acrodef{TDL}{tapped delay line}
\acrodef{OMP}{orthogonal matching pursuit}
\acrodef{DCS-SOMP}{distributed compressed sensing-simultaneous orthogonal matching pursuit}
\acrodef{DCS}{distributed compressed sensing}
\acrodef{CS}{compressed sensing}
\acrodef{CoSOMP}{compressive sampling matched pursuit}
\acrodef{SOMP}{simultaneous OMP}
\acrodef{RA-ORMP}{rank-aware order recursive matching pursuit}
\acrodef{G-BPDN}{group basis pursuit denoising}
\acrodef{GCS}{group sparse compressed sensing}
\acrodef{MMV}{multiple measurement vectors}
\acrodef{SMV}{single measurement vector}
\acrodef{ReMBo}{reduce MMV and boost}
\acrodef{MLE}{maximum likelihood estimation}
\acrodef{IQML}{iterative quadratic maximum likelihood}
\acrodef{RMSE}{root-mean-square error}
\acrodef{LS}{least squares}
\acrodef{RSE}{root-square error}
\acrodef{RMS}{root-mean-square}
\acrodef{MMSE}{minimum mean square error}
\acrodef{EM}{expectation maximization}
\acrodef{SAGE}{space-alternating generalized expectation maximization}
\acrodef{OFDM}{orthogonal frequency division multiplexing}
\begin{document}                        % End of preamble, start of text.
\title{Position and Orientation Estimation through Millimeter Wave MIMO in 5G Systems}
\author{Arash Shahmansoori, Gabriel E. Garcia, Giuseppe Destino,~\IEEEmembership{Member,~IEEE}, Gonzalo Seco-Granados,~\IEEEmembership{Senior Member,~IEEE}, and Henk Wymeersch,~\IEEEmembership{Member,~IEEE}
\thanks{
Arash Shahmansoori and Gonzalo Seco-Granados are with the Department of Telecommunications and Systems Engineering, Universitat Aut\`{o}noma de Barcelona, Spain, email: \{arash.shahmansoori,gonzalo.seco\}@uab.cat. Gabriel E. Garcia and Henk Wymeersch are with the Department of Signals and Systems, Chalmers University of Technology, Sweden, email: \{ggarcia,henkw\}@chalmers.se. Giuseppe Destino is with the center for wireless communications, University of Oulu, Finland, email:  giuseppe.destino@oulu.fi. This work was financially supported by EU FP7 Marie Curie Initial Training Network MULTI-POS (Multi-technology Positioning Professionals) under grant nr. 316528, the EU-H2020 project HIGHTS (High Precision Positioning for Cooperative ITS Applications) under grant nr. MG-3.5a-2014-636537, the VINNOVA COPPLAR project, funded under Strategic Vehicle Research and Innovation grant nr. 2015-04849, and R\&D Project of Spanish Ministry of Economy and Competitiveness TEC2014-53656-R. Part of this work was previously presented at the 2015 IEEE Global Telecommunications (GLOBECOM) Conference \cite{Arash}.}
}
%}
\maketitle
\begin{abstract}
Millimeter wave signals and large antenna arrays are considered enabling technologies for future 5G networks. While their benefits for achieving high-data rate communications are well-known, their potential advantages for accurate positioning are largely undiscovered. We derive the \ac{CRB} on position and rotation angle estimation uncertainty from millimeter wave signals from a single transmitter, in the presence of scatterers. We also present a novel two-stage algorithm for position and rotation angle estimation that attains the \ac{CRB} for average to high signal-to-noise ratio. The algorithm is based on multiple measurement vectors matching pursuit for coarse estimation, followed by a refinement stage based on the space-alternating generalized expectation maximization algorithm. 
 We find that  accurate position and rotation angle estimation is possible using signals from a single transmitter, in either line-of-sight, non-line-of-sight, or obstructed-line-of-sight conditions. 
\end{abstract}
\section{Introduction}
Fifth generation (5G) communication networks will likely adopt \ac{mm-wave} and massive \ac{MIMO}  technologies, thanks to a number of favorable properties. In particular, operating at carrier frequencies beyond 30 GHz, with large available bandwidths, \ac{mm-wave} can provide extremely high data rates to users through dense spatial multiplexing by using a large number of antennas \cite{Zhouyue,Rappaport}. While these properties are desirable for 5G services, \ac{mm-wave} communications also face a number of challenges. Among these, the severe path loss at high carrier frequencies stands out. The resulting loss in \ac{SNR} must be compensated through sophisticated beamforming at the transmitter and/or receiver side, leading to highly directional links \cite{Wang,Hur,Tsang}. However, beamforming requires knowledge of the propagation channel. Significant progress has been made in \ac{mm-wave} channel estimation, by exploiting sparsity and related compressed sensing tools, such as \ac{DCS-SOMP} \cite{Duarte}, \ac{CoSOMP} \cite{Duarte2}, and \ac{GCS} \cite{Bolcskei}.  In particular, since at \ac{mm-wave} frequencies only the \ac{LOS} path and a few dominant multi-path components contribute to the received power, 
\ac{mm-wave} channels are sparse in the angular domain \cite{BspaceSayeed,widebandbrady}.  This is because in \ac{mm-wave} frequencies, the received power of diffuse scattering and multiple-bounce specular reflections are much lower than that in \ac{LOS} and single-bounce specular reflection \cite{Martinez-Ingles,Vaughan,mmMAGIC}. Different \ac{CS} methods for \ac{mm-wave} channel estimation are proposed in\cite{Marzi,LeeJ,AlkhateebA,ChoiJ,AlkhateebC,HanY,LeeJ2,Ramasamy,BerrakiD}. In \cite{Marzi}, a method for the estimation of \ac{AOA}, \ac{AOD}, and channel gains is proposed based on the  compressive beacons on the downlink. 
A method for the continuous estimation of \ac{mm-wave} channel parameters is proposed in \cite{Ramasamy}, while \cite{LeeJ2} applies CS tools with refinement in the angular domain. In \cite{LeeJ},  a CS method is proposed based on the redundant dictionary matrices.
A two-stage algorithm with one-time feedback that is robust to noise is used in \cite{HanY}. In \cite{AlkhateebA}, an adaptive \ac{CS} method is proposed based on a hierarchical multi-resolution codebook design for the estimation of single-path and multi-path \ac{mm-wave} channels. In \cite{ChoiJ}, a beam selection procedure for the multiuser \ac{mm-wave} \ac{MIMO} channels with analog beamformers is proposed. 
In \cite{AlkhateebC}, a  \ac{CS} approach with reduced training overhead was considered. 
Finally, \ac{CS} tools are used in \cite{BerrakiD} for the sparse estimation of power angle profiles of the \ac{mm-wave} channels and compared with the codebook designs in terms of overhead reduction. However, in all the aforementioned papers, a narrow-band \ac{mm-wave} channel model is used. When the bandwidths becomes larger, one needs to consider the effect of the delays of different paths in the \ac{mm-wave} channel model, i.e., the wide-band \ac{mm-wave} channel model. 

Channel estimation provides information of the \ac{AOA}/\ac{AOD} and thus of the relative location of the transmitter and receiver. In addition, location information can serve as a proxy for channel information to perform beamforming: when the location of the user is known, the \ac{BS} can steer its transmission to the user, either directly or through a reflected path. This leads to synergies between localization and communication. The use of 5G technologies to obtain position and orientation was previously explored in \cite{sanchis2002novel,DenSaya,vari2014mmwaves} for mm-wave and in \cite{hu2014esprit,Dardari,savic2015fingerprinting} for massive MIMO. The early work \cite{sanchis2002novel} considered estimation and tracking of \ac{AOA} through beam-switching. User localization was treated in \cite{DenSaya}, formulated as a hypothesis testing problem, limiting the spatial resolution. A different approach was taken in \cite{vari2014mmwaves}, where meter-level positioning accuracy was obtained by measuring received signal strength levels. A
location-aided beamforming method was proposed in \cite{NGarcia} to speed up initial access between nodes. In the massive \ac{MIMO} case, \cite{hu2014esprit} considered the estimation of angles, \cite{NGarcia2} proposed a direct localization method by jointly processing the observations at the distributed
massive \ac{MIMO} \ac{BS}s, while \cite{Dardari} treated the joint estimation of delay, \ac{AOD}, and \ac{AOA}, in the \ac{LOS} conditions and evaluated the impact of errors in delays and phase shifters, and \cite{Arash} derived sufficient conditions for a nonsingular \ac{FIM} of delay, \ac{AOD}, \ac{AOA}, and channel coefficients. A hybrid \ac{TDOA}, \ac{AOA}, and \ac{AOD} localization was proposed in \cite{linhyb} using linearization. In \cite{savic2015fingerprinting}, positioning was solved using a Gaussian process regressor, operating on a vector of received signal strengths through fingerprinting. While latter this approach is able to exploit \ac{NLOS} propagation, it does not directly harness the geometry of the environment. Complementarily to the use of \ac{mm-wave} frequencies, approaches for localization using \ac{cm-wave} signals have been recently proposed as well. The combination of \ac{TDOA}s and \ac{AOA}s using an extended Kalman filter (EKF) was presented in \cite{DBLP:journals/twc/KoivistoCWHTLKV17,DBLP:journals/corr/KoivistoHCKLV16}, where the \ac{MS} has a single antenna, while the \ac{BS} employs an antenna array. This method assumes \ac{LOS} propagation thanks to the high density of access nodes and provides sub-meter accuracy even for moving devices.

In this paper, we show that \ac{mm-wave} and large \ac{MIMO} are enabling technologies for accurate positioning and device orientation estimation with only one \ac{BS}, even when the \ac{LOS} path is blocked. The limited scattering and high-directivity are unique characteristics of the \ac{mm-wave} channel and large \ac{MIMO} systems, respectively. We derive fundamental bounds on the position and orientation estimation accuracy, for \ac{LOS}\footnote{\ac{LOS} is defined as the condition where the \ac{LOS} path exists and there are no scatterers.}, \ac{NLOS}\footnote{\ac{NLOS} is defined as the condition where there are scatterers and the LOS path is not blocked.}, and \ac{OLOS}\footnote{\ac{OLOS} is referred to the condition where the LOS path is blocked and only the signals from the scatterers are received.} conditions. These bounds indicate that the information from the \ac{NLOS} links help to estimate the location and orientation of the \ac{MS}. We also propose a novel three-stage position and orientation estimation technique, which is able to attain the bounds at average to high \ac{SNR}. The first stage of the technique harnesses sparsity of the \ac{mm-wave} channel in the \ac{AOA} and \ac{AOD} domain \cite{BspaceSayeed,widebandbrady}. Moreover, the sparsity support does not vary significantly with frequency, allowing us to use \ac{DCS-SOMP} across different carriers. The delay can then be estimated on a per-path basis. As \ac{DCS-SOMP} limits the \ac{AOA} and \ac{AOD} to a predefined grid, we propose a refinement stage, based on the \ac{SAGE} algorithm. Finally, in the last stage, we employ a least-squares approach with \ac{EXIP} to recover position and orientation \cite{Stoicapp,Swindlehurstt}. 

\begin{figure}   
\psfrag{x}{\small $x$}
\psfrag{y}{\small  $y$}
\psfrag{d0}{\small $d_{0}$}
\psfrag{dk1}{\small  $d_{k,1}$}
\psfrag{dk2}{\small $d_{k,2}$}
\psfrag{dk}{\small $d_{k}=d_{k,1}+d_{k,2}$}
\psfrag{sk}{\hspace{-1mm} $\mathbf{s}_{k}$}
\psfrag{tt0}{\small \hspace{-1mm} $\theta_{\mathrm{Tx},0}$}
\psfrag{tt1}{\small  \hspace{-4mm} $\theta_{\mathrm{Tx},k}$}
\psfrag{tt1b}{\small  \hspace{-8mm} $\pi-(\theta_{\mathrm{Rx},k}+\alpha)$}
\psfrag{rr0}{\small  \hspace{-8mm} $\pi-\theta_{\mathrm{Rx},0}$}
\psfrag{rr1}{\small \hspace{-8mm} $\pi-\theta_{\mathrm{Rx},k}$}
\psfrag{rr1b}{\small \hspace{-8mm} $\pi-(\theta_{\mathrm{Rx},k}+\alpha)$}
\psfrag{alphab}{\small  \hspace{-4mm} $\alpha$}
\psfrag{q}{ \hspace{-1mm} $\mathbf{q}$}
\psfrag{qk}{ \hspace{-1mm} $\widetilde{\mathbf{q}}_{k}$}
\psfrag{p}{  \hspace{-4mm} $\mathbf{p}$}
\psfrag{BS}[][c]{BS}
\psfrag{VBS}[][c]{virtual BS}
\psfrag{MS}[][c]{MS}
\centering
\includegraphics[width=0.9\columnwidth]{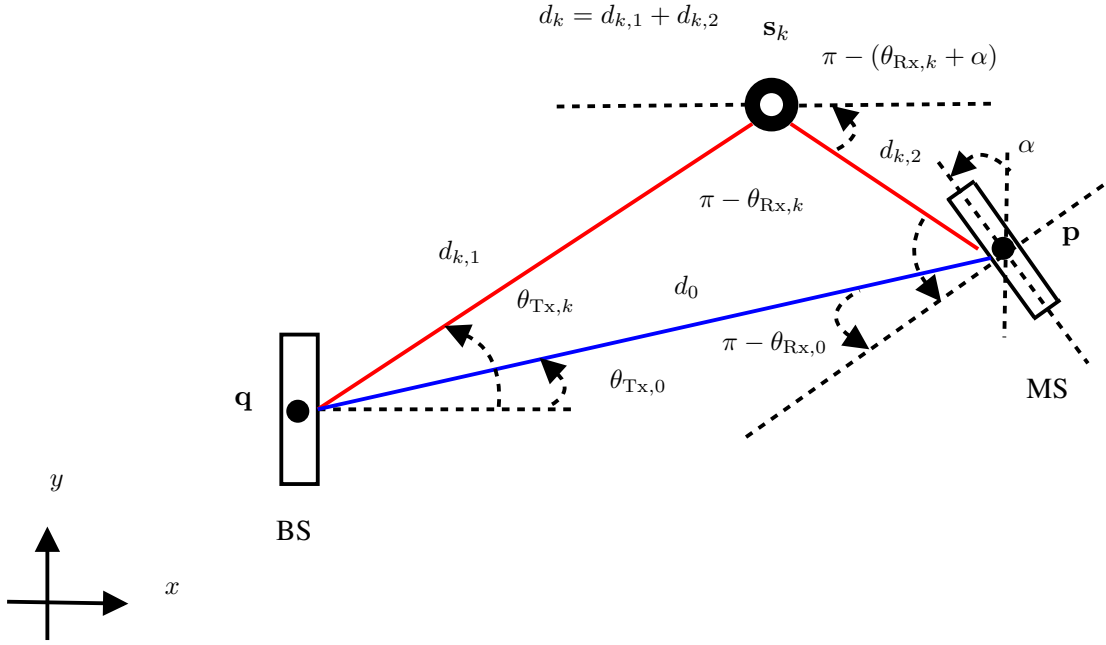}
  \caption{Two dimensional illustration of the \ac{LOS} (blue link) and \ac{NLOS} (red link) based positioning problem. The \ac{BS} location $\mathbf{q}$ and \ac{BS} orientation are known, but arbitrary. The location of the \ac{MS} $\mathbf{p}$, scatterer $\mathbf{s}_{k}$, rotation angle $\alpha$, \ac{AOA}s $\{\theta_{\textup{Rx},k}\}$, \ac{AOD}s $\{\theta_{\textup{Tx},k}\}$, the channels between \ac{BS}, \ac{MS}, and scatterers, and the distance between the antenna centers are unknown.}
  \label{NLOS_Link}
\end{figure}

\section{System Model}\label{SEC:Formulation}
We consider a \ac{MIMO} system with a \ac{BS} equipped with $N_{t}$ antennas and a \ac{MS} equipped by $N_{r}$ antennas operating at a carrier frequency $f_c$ (corresponding to wavelength $\lambda_c$) and bandwidth $B$. 
Locations of the \ac{MS} and \ac{BS} are denoted by $\mathbf{p}=[p_{x}, p_{y}]^{\mathrm{T}}\in\mathbb{R}^{2}$ and $\mathbf{q}=[q_{x}, q_{y}]^{\mathrm{T}}\in\mathbb{R}^{2}$ with the $\alpha\in[0, 2\pi)$ denoting the rotation angle of the \ac{MS}'s antenna array. The value of $\mathbf{q}$ is assumed to be known, while $\mathbf{p}$ and $\alpha$ are unknown. 
\subsection{Transmitter Model}
We consider the transmission of \ac{OFDM} signals as in \cite{khateeb3}, where a \ac{BS} with hybrid analog/digital precoder communicates with a single \ac{MS}. At the \ac{BS}, $G$ signals are transmitted sequentially, where the $g$-th transmission comprises  $M_{t}$ simultaneously transmitted symbols $\mathbf{x}^{(g)}[n]=[x_{1}[n],\ldots,x_{M_{t}}[n]]^{\mathrm{T}} \in \mathbb{C}^{M_{t}}$ for each subcarrier $n=0,\ldots,N-1$. The symbols  
are first precoded and then transformed to the time-domain using $N$-point \acf{IFFT}. A \acf{CP} of length $T_{\mathrm{CP}}=DT_{s}$ is added before applying the \ac{RF} precoding where $D$ is the length of \ac{CP} in symbols. Here, $T_{s}=1/B$ denotes the sampling period and  $T_{\mathrm{CP}}$ is assumed to exceed the delay spread of the channel. The transmitted signal over subcarrier $n$ at time $g$ can be expressed as $\mathbf{F}^{(g)}[n]\mathbf{x}^{(g)}[n]$. The beamforming matrix $\mathbf{F}[n] \in \mathbb{C}^{N_{t}\times M_{t}}$ is defined as $\mathbf{F}[n]=\mathbf{F}_{\mathrm{RF}}\mathbf{F}_{\mathrm{BB}}[n]$ where $\mathbf{F}_{\mathrm{RF}}$ is implemented using the analog phase shifters with the entries of the form $e^{j\phi_{m,n}}$, where $\{\phi_{m,n}\}$ are given phases, and $\mathbf{F}_{\mathrm{BB}}[n]$ is the digital beamformer,  and overall they satisfy a total power constraint $\Vert\mathbf{F}_{\mathrm{RF}}\mathbf{F}_{\mathrm{BB}}[n]\Vert_{\mathrm{F}}=1$. %or the unitary power constraint $\mathbf{F}_{\mathrm{RF}}\mathbf{F}_{\mathrm{BB}}[n]\in\mathcal{U}_{N_{t}\times M_{t}}$ where $\mathcal{U}_{N_{t}\times M_{t}}$ denotes the set of semi-unitary matrices \cite{khateeb3}. 
Considering the sparsity of the \ac{mm-wave} channels one usually needs much less beams $M_{t}$ than antenna elements $N_{t}$, i.e., $M_{t} \ll N_{t}$. Also, the presence of $\mathbf{F}[n]$ in the proposed model leads to the extension of system model to multi-user mm-wave downlink systems with a limited feedback channel from \ac{MS}s to the \ac{BS}. 
%\textcolor{red}{Our work does not assume any specific beamformer, though precoding towards a specific user will not be helpful for positioning other user. Our approach is thus more compatible with beam reference signal (initial access) procedures, and can be complemented with a Bayesian recursive tracker with user-specific precoding. This latter aspect is out of the scope of the current work.}
Our work does not assume any specific beamformer. We will provide general expressions that permit the study of the impact on performance and optimization of different choices of beamformers $\mathbf{F}^{(g)}[n]$ and signals $\mathbf{x}^{(g)}[n]$, although this is out of the scope of the paper. Our approach is also  compatible with beam reference signal (initial access) procedures, and it could be complemented with a Bayesian recursive tracker with user-specific precoding.
 
\subsection{Channel Model}
Fig.~\ref{NLOS_Link} shows the position-related parameters of the channel. These parameters include 
 $\theta_{\mathrm{Rx},k}$, $\theta_{\mathrm{Tx},k}$, and $d_{k}=c\tau_{k}$,  denoting the \ac{AOA}, \ac{AOD}, and the path length (with \ac{TOA} $\tau_{k}$ and the speed of light $c$) of the $k$-th path ($k=0$ for the \ac{LOS} path and  $k>0$ the \ac{NLOS}  paths). For each NLOS path, there is a scatterer with unknown location $\mathbf{s}_{k}$, for which we define  $d_{k,1}=\Vert\mathbf{s}_{k}-\mathbf{q}\Vert_{2}$ and $d_{k,2}=\Vert\mathbf{p}-\mathbf{s}_{k}\Vert_{2}$. We now introduce the channel model, under a frequency-dependent array response \cite{widebandbrady}, suitable for wideband communication (with fractional bandwidth  $B/f_c$ up to $50\%$). Assuming $K+1$ paths and a channel that remains constant during the transmission of $G$ symbols, the $N_r \times N_t$ channel matrix associated with subcarrier $n$ is expressed as
 \begin{equation}\label{Channel1}
 \mathbf{H}[n]=\mathbf{A}_{\mathrm{Rx}}[n]\mathbf{\Gamma}[n]\mathbf{A}^{\mathrm{H}}_{\mathrm{Tx}}[n],
 \end{equation}
for  response vectors
 \begin{align}
 \mathbf{A}_{\mathrm{Tx}}[n]& =[\mathbf{a}_{\mathrm{Tx},n}(\theta_{\mathrm{Tx},0}),\ldots,\mathbf{a}_{\mathrm{Tx},n}(\theta_{\mathrm{Tx},K})], \\
 \mathbf{A}_{\mathrm{Rx}}[n]& =[\mathbf{a}_{\mathrm{Rx},n}(\theta_{\mathrm{Rx},0}),\ldots,\mathbf{a}_{\mathrm{Rx},n}(\theta_{\mathrm{Rx},K})],
 \end{align}
 and
 \begin{align}
\mathbf{\Gamma}[n]=\sqrt{N_{t}N_{r}}\mathrm{diag}\left\{ \frac{h_{0}}{\sqrt{\rho_0}}e^{-j2\pi n\tau_{0}/(NT_{s})},\ldots,\frac{h_{K}}{\sqrt{\rho_K}}e^{-j2\pi n\tau_{K}/(NT_{s})}\right\},
 \end{align}
for path loss $\rho_k$ and complex channel gain $h_k$, respectively, of the $k$-th path. For later use, we introduce $\tilde{h}_{k}=\sqrt{(N_{t}N_{r})/\rho_{k}}h_{k}$ and $\gamma_{n}(h_{k},\tau_{k})=\tilde{h}_{k}e^{-j2\pi n\tau_{k}/(NT_{s})}$. 

The structure of the frequency-dependent antenna steering and response vectors $\mathbf{a}_{\mathrm{Tx},n}(\theta_{\mathrm{Tx},k})\in \mathbb{C}^{N_t}$ and $\mathbf{a}_{\mathrm{Rx},n}(\theta_{\mathrm{Rx},k})\in \mathbb{C}^{N_r}$ depends on the specific array structure. For the case of a \ac{ULA}, which will be the example studied in this paper, we recall that (the response vector $\mathbf{a}_{\mathrm{Rx},n}(\theta_{\mathrm{Rx},k})$ is obtained similarly)
 \begin{align}\label{steringvector1}
& \mathbf{a}_{\mathrm{Tx},n}(\theta_{\mathrm{Tx},k}) =\\ 
%&\frac{1}{\sqrt{N_{t}}}[
%e^{j\frac{2\pi}{\lambda_{n}}d\sin(\theta_{\mathrm{Tx},k})},\ldots,e^{j(N_{t}-1)\frac{2\pi}{\lambda_{n}}d\sin(\theta_{\mathrm{Tx},k})}
%]^{\mathrm{T}}, \nonumber
&\frac{1}{\sqrt{N_{t}}}[
e^{-j \frac{N_{t}-1}{2}\frac{2\pi}{\lambda_{n}}d\sin(\theta_{\mathrm{Tx},k})},\ldots,e^{j\frac{N_{t}-1}{2}\frac{2\pi}{\lambda_{n}}d\sin(\theta_{\mathrm{Tx},k})}
]^{\mathrm{T}}, \nonumber
\end{align}
where $\lambda_{n} = c/(n/(NT_s)+f_{c})$ is the signal wavelength at the $n$-th subcarrier and $d$ denotes the distance between the antenna elements (we will use $d=\lambda_{c}/2$). We note that when $B\ll f_c$, $\lambda_{n} \approx \lambda_c$, and $\eqref{steringvector1}$ reverts to the standard narrow-band model. 
\subsection{Received Signal Model}
The received signal for subcarrier $n$ and transmission $g$, after \ac{CP} removal and \acf{FFT}, can be expressed as
\begin{equation}
\mathbf{y}^{(g)}[n]=\mathbf{H}[n]\mathbf{F}^{(g)}[n]\mathbf{x}^{(g)}[n]+\mathbf{n}^{(g)}[n],\label{Receivedb1}
\end{equation}
where  $\mathbf{n}^{(g)}[n]\in\mathbb{C}^{N_r}$ is a Gaussian noise vector with zero mean and variance $N_{0}/2$ per real dimension. Our goal is now to estimate the position $\mathbf{p}$ and orientation $\alpha$ of the \ac{MS} from $\{\mathbf{y}^{(g)}[n]\}_{\forall n, g}$. We will first derive a fundamental lower bound on the estimation uncertainty and then propose a novel practical estimator.

\section{Position and Orientation Estimation: Fundamental Bounds}\label{SEC:FundamentalBound}
In this section, we  derive  the \ac{FIM} and the \acf{CRB} for the estimation problem of position and orientation of the \ac{MS} for  \ac{LOS}, \ac{NLOS}, and \ac{OLOS}. To simplify the notation and without loss of generality, we consider the case of $G=1$, i.e., only 1 OFDM symbol is transmitted. 
\subsection{FIM Derivation for Channel Parameters}
Let $\boldsymbol{\eta}\in\mathbb{R}^{5(K+1)}$ be the vector consisting of the unknown channel parameters
\begin{equation}\label{Parameters1}
\boldsymbol{\eta}=\begin{bmatrix}\boldsymbol{\eta}^{\mathrm{T}}_{0},\ldots,\boldsymbol{\eta}^{\mathrm{T}}_{K}\end{bmatrix}^{\mathrm{T}},
\end{equation}
in which $\boldsymbol{\eta}_{k}$ consists of the unknown channel parameters (delay, \ac{AOD}, \ac{AOA}, and channel coefficients) for the $k$-th path
\begin{equation}\label{Parameters2}
\boldsymbol{\eta}_{k}=\begin{bmatrix}
\tau_{k},\boldsymbol{\theta}_{k}^{\mathrm{T}},\tilde{\mathbf{h}}_{k}^{\mathrm{T}}
\end{bmatrix}^{\mathrm{T}},
\end{equation}
where $\tilde{\mathbf{h}}_{k}=[\tilde{h}_{\mathrm{R},k},\tilde{h}_{\mathrm{I},k}]^{\mathrm{T}}$ contains the real and imaginary parts defined as $\tilde{h}_{\mathrm{R},k}$ and $\tilde{h}_{\mathrm{I},k}$, respectively, and $\boldsymbol{\theta}_{k}=\begin{bmatrix}\theta_{\mathrm{Tx},k},\theta_{\mathrm{Rx},k}\end{bmatrix}^{\mathrm{T}}$. 

Defining $\hat{\boldsymbol{\eta}}$ as the unbiased estimator of $\boldsymbol{\eta}$, the mean squared error (MSE) is bounded as \cite{Kay}
\begin{equation}\label{Parameters3}
\mathbb{E}_{\mathbf{y}\vert\boldsymbol{\eta}}\left[(\hat{\boldsymbol{\eta}}-\boldsymbol{\eta})(\hat{\boldsymbol{\eta}}-\boldsymbol{\eta})^{\mathrm{T}}\right]\succeq\mathbf{J}^{-1}_{\boldsymbol{\eta}},
\end{equation}
in which $\mathbb{E}_{\mathbf{y}\vert\boldsymbol{\eta}}[.]$ denotes the expectation parameterized by the unknown parameters $\boldsymbol{\eta}$, and $\mathbf{J}_{\boldsymbol{\eta}}$ is the $5(K+1)\times 5(K+1)$ FIM defined as
\begin{equation}\label{Parameters4}
\mathbf{J}_{\boldsymbol{\eta}}\triangleq\mathbb{E}_{\mathbf{y}\vert\boldsymbol{\eta}}\left[-\frac{\partial^{2} \ln f(\mathbf{y}\vert\boldsymbol{\eta})}{\partial\boldsymbol{\eta}\partial\boldsymbol{\eta}^{T}}\right],
\end{equation}
where $f(\mathbf{y}\vert\boldsymbol{\eta})$ is the likelihood function of the random vector $\mathbf{y}$ conditioned on $\boldsymbol{\eta}$. More specifically,  $f(\mathbf{y}\vert\boldsymbol{\eta})$ can be written as \cite{Poor}
\begin{equation}\label{Parameters5b}
\!\!f(\mathbf{y}\vert\boldsymbol{\eta})\!\propto\! \exp\!\left\{\!\frac{2}{N_{0}}\!\!\sum^{N-1}_{n=0}\Re\{\boldsymbol{\mu}^{\mathrm{H}}[n]\mathbf{y}[n]\}\!-\!\frac{1}{N_{0}}\!\!\sum^{N-1}_{n=0}\Vert\boldsymbol{\mu}[n]\Vert_{2}^{2}\!\right\}\!,
\end{equation}
where $\boldsymbol{\mu}[n]\triangleq\mathbf{H}[n]\mathbf{F}[n]\mathbf{x}[n]$ and $\propto$ denotes equality up to irrelevant constants. 

The \ac{FIM} in \eqref{Parameters4} can be structured as
\begin{equation}\label{Parameters6w}
\mathbf{J}_{\boldsymbol{\eta}}=\begin{bmatrix}
\mathbf{\Psi}(\boldsymbol{\eta}_{0},\boldsymbol{\eta}_{0})&\ldots&\mathbf{\Psi}(\boldsymbol{\eta}_{0},\boldsymbol{\eta}_{K})\\
\vdots&\ddots&\vdots\\
\mathbf{\Psi}(\boldsymbol{\eta}_{K},\boldsymbol{\eta}_{0})&\ldots&\mathbf{\Psi}(\boldsymbol{\eta}_{K},\boldsymbol{\eta}_{K})
\end{bmatrix},
\end{equation}
in which  $\mathbf{\Psi}(\mathbf{x}_{r},\mathbf{x}_{s})$ is defined as
\begin{equation}\label{Parameters6ww}
\mathbf{\Psi}(\mathbf{x}_{r},\mathbf{x}_{s})\triangleq\mathbb{E}_{\mathbf{y}\vert\boldsymbol{\eta}}\left[-\frac{\partial^{2} \ln f(\mathbf{y}\vert\boldsymbol{\eta})}{\partial \mathbf{x}_{r}\partial \mathbf{x}^{\mathrm{T}}_{s}}\right].
\end{equation}
The $5 \times 5$ matrix $\mathbf{\Psi}(\boldsymbol{\eta}_{r},\boldsymbol{\eta}_{s})$ is structured as 
\begin{equation}\label{Parameters8w}
\mathbf{\Psi}(\boldsymbol{\eta}_{r},\boldsymbol{\eta}_{s})=\begin{bmatrix}
\Psi(\tau_{r},\tau_{s})&\mathbf{\Psi}(\tau_{r},\boldsymbol{\theta}_{s})&\mathbf{\Psi}(\tau_{r},\mathbf{h}_{s})\\
\mathbf{\Psi}(\boldsymbol{\theta}_{r},\tau_{s})&\mathbf{\Psi}(\boldsymbol{\theta}_{r},\boldsymbol{\theta}_{s})&\mathbf{\Psi}(\boldsymbol{\theta}_{r},\mathbf{h}_{s})\\
\mathbf{\Psi}(\mathbf{h}_{r},\tau_{s})&\mathbf{\Psi}(\mathbf{h}_{r},\boldsymbol{\theta}_{s})&\mathbf{\Psi}(\mathbf{h}_{r},\mathbf{h}_{s})
\end{bmatrix}.
\end{equation}
The entries of $\mathbf{\Psi}(\boldsymbol{\eta}_{r},\boldsymbol{\eta}_{s})$ are derived in Appendix \ref{elements}.

\subsection{FIM for Position and Orientation}\label{Trans_Convert}
We determine the FIM in the position space through a transformation of variables from $\boldsymbol{\eta}$ to $\tilde{\boldsymbol{\eta}}=\begin{bmatrix}\tilde{\boldsymbol{\eta}}^{\mathrm{T}}_{0},\ldots,\tilde{\boldsymbol{\eta}}^{\mathrm{T}}_{K}\end{bmatrix}^{\mathrm{T}}$, where $\tilde{\boldsymbol{\eta}}_{k}=\begin{bmatrix}\mathbf{s}^{\mathrm{T}}_{k},\tilde{\mathbf{h}}^{\mathrm{T}}_{k}\end{bmatrix}^{\mathrm{T}}$  for $k > 0$ and $\tilde{\boldsymbol{\eta}}_{0}=\begin{bmatrix}\mathbf{p}^{\mathrm{T}},\alpha,\tilde{\mathbf{h}}^{\mathrm{T}}_{0}\end{bmatrix}^{\mathrm{T}}$. If the \ac{LOS} path is blocked (i.e., \ac{OLOS}), we note that we must consider $\boldsymbol{\eta}_{\mathrm{olos}}=[\boldsymbol{\eta}^{\mathrm{T}}_{1},\ldots,\boldsymbol{\eta}^{\mathrm{T}}_{K}]^{\mathrm{T}}$ and $\tilde{\boldsymbol{\eta}}_{\mathrm{olos}}=[\mathbf{p}^{\mathrm{T}},\alpha,\tilde{\boldsymbol{\eta}}^{\mathrm{T}}_{1},\ldots,\tilde{\boldsymbol{\eta}}^{\mathrm{T}}_{K}]^{\mathrm{T}}$.

The \ac{FIM} of $\tilde{\boldsymbol{\eta}}$ is obtained by means of the $(4K+5)\times 5(K+1)$ transformation matrix $\mathbf{T}$ as
\begin{equation}\label{TFIM1}
\mathbf{J}_{\tilde{\boldsymbol{\eta}}}=\mathbf{T}\mathbf{J}_{\boldsymbol{\eta}}\mathbf{T}^{\mathrm{T}},
\end{equation}
where 
\begin{equation}\label{TFIM2}
\mathbf{T}\triangleq\frac{\partial\boldsymbol{\eta}^{\mathrm{T}}}{\partial\tilde{\boldsymbol{\eta}}}.
\end{equation}
The entries of $\mathbf{T}$ can be obtained by the relations between the parameters in $\boldsymbol{\eta}$ and $\tilde{\boldsymbol{\eta}}$ from the geometry of the problem shown in Fig. \ref{NLOS_Link} as: 
\begin{align}
\tau_{0} & = \Vert\mathbf{p}-\mathbf{q}\Vert_{2}/c, \label{TFIM2xy1}\\
\tau_{k}& = \Vert\mathbf{q}-\mathbf{s}_{k}\Vert_{2}/c+\Vert\mathbf{p}-\mathbf{s}_{k}\Vert_{2}/c,\: k>0 \label{TFIM2xy1b}\\
\theta_{\mathrm{Tx},0} & = \arccos((p_{x}-q_{x})/\Vert\mathbf{p}-\mathbf{q}\Vert_{2}),\label{TFIM2xy2}\\
\theta_{\mathrm{Tx},k} & = \arccos((s_{k,x}-q_{x})/\Vert\mathbf{s}_{k}-\mathbf{q}\Vert_{2}),\: k>0\label{TFIM2xy3}\\
\theta_{\mathrm{Rx},k} & = \pi -\arccos((p_{x}-s_{k,x})/\Vert\mathbf{p}-\mathbf{s}_{k}\Vert_{2})-\alpha ,\: k>0\label{TFIM2xy4}\\
\theta_{\mathrm{Rx},0} & =\pi+\arccos((p_{x}-q_{x})/\Vert\mathbf{p}-\mathbf{q}\Vert_{2})-\alpha.\label{TFIM2xy5}
%\cos(\pi-(\theta_{\mathrm{Rx},k}+\alpha))& = (p_{x}-s_{k,x})/\Vert\mathbf{p}-\mathbf{s}_{k}\Vert_{2},\: k>0\label{TFIM2xy4}\\
%\alpha& =\pi+\arccos((p_{x}-q_{x})/\Vert\mathbf{p}-\mathbf{q}\Vert_{2})-\theta_{\mathrm{Rx},0},\label{TFIM2xy5}\\
\end{align}
Consequently, we obtain
\begin{equation}\label{TFIM3}
\mathbf{T}=\begin{bmatrix}
\mathbf{T}_{0,0}&\ldots&\mathbf{T}_{K,0}\\
\vdots&\ddots&\vdots\\
\mathbf{T}_{0,K}&\ldots&\mathbf{T}_{K,K}
\end{bmatrix},
\end{equation}
in which $\mathbf{T}_{k,k'}$ is defined as
\begin{equation}\label{TFIM4}
\mathbf{T}_{k,k'}\triangleq\frac{\partial\boldsymbol{\eta}^{\mathrm{T}}_{k}}{\partial\tilde{\boldsymbol{\eta}}_{k'}}.
\end{equation}
For $k'\neq 0$, $\mathbf{T}_{k,k'}$ is obtained as
\begin{equation}\label{TFIM5}
\mathbf{T}_{k,k'}=\begin{bmatrix}
\partial\tau_{k}/\partial\mathbf{s}_{k'}&\partial\boldsymbol{\theta}^{\mathrm{T}}_{k}/
\partial\mathbf{s}_{k'}&\partial\tilde{\mathbf{h}}^{\mathrm{T}}_{k}/
\partial\mathbf{s}_{k'}\\
\partial\tau_{k}/\partial\tilde{\mathbf{h}}_{k'}&\partial\boldsymbol{\theta}^{\mathrm{T}}_{k}/
\partial\tilde{\mathbf{h}}_{k'}&\partial\tilde{\mathbf{h}}^{\mathrm{T}}_{k}/\partial\tilde{\mathbf{h}}_{k'}
\end{bmatrix},
\end{equation}
and $\mathbf{T}_{k,0}$ is obtained as
\begin{equation}\label{TFIM6}
\mathbf{T}_{k,0}=\begin{bmatrix}
\partial\tau_{k}/\partial\mathbf{p}&\partial\boldsymbol{\theta}^{\mathrm{T}}_{k}/
\partial\mathbf{p}&\partial\tilde{\mathbf{h}}^{\mathrm{T}}_{k}/
\partial\mathbf{p}\\
\partial\tau_{k}/
\partial\alpha&\partial\boldsymbol{\theta}^{\mathrm{T}}_{k}/
\partial\alpha&\partial\tilde{\mathbf{h}}^{\mathrm{T}}_{k}/
\partial\alpha\\
\partial\tau_{k}/\partial\tilde{\mathbf{h}}_{0}&\partial\boldsymbol{\theta}^{\mathrm{T}}_{k}/
\partial\tilde{\mathbf{h}}_{0}&\partial\tilde{\mathbf{h}}^{\mathrm{T}}_{k}/\partial\tilde{\mathbf{h}}_{0}
\end{bmatrix},
\end{equation}
where 
\begin{align*}
\partial\tau_{0}/\partial\mathbf{p} & =\frac{1}{c}\begin{bmatrix}\cos(\theta_{\mathrm{Tx},0}),\sin(\theta_{\mathrm{Tx},0})\end{bmatrix}^{\mathrm{T}} ,\\
\partial\theta_{\mathrm{Tx},0}/\partial\mathbf{p}  & =\frac{1}{\Vert\mathbf{p}-\mathbf{q}\Vert_{2}}\begin{bmatrix}-\sin(\theta_{\mathrm{Tx},0}), \cos(\theta_{\mathrm{Tx},0})\end{bmatrix}^{\mathrm{T}},\\
\partial\theta_{\mathrm{Rx},0}/\partial\mathbf{p}  & =\frac{1}{\Vert\mathbf{p}-\mathbf{q}\Vert_{2}}\begin{bmatrix}-\sin(\theta_{\mathrm{Tx},0}), \cos(\theta_{\mathrm{Tx},0})\end{bmatrix}^{\mathrm{T}},\\
\partial\theta_{\mathrm{Rx},k}/\partial\alpha  & =-1, k\ge 0\\
\end{align*}
\begin{align*}
\partial\tau_{k}/\partial\mathbf{p} & = \frac{1}{c}\begin{bmatrix}\cos(\pi-\theta_{\mathrm{Rx},k}),
-\sin(\pi-\theta_{\mathrm{Rx},k})\end{bmatrix}^{\mathrm{T}},\:k> 0\\
\partial\tau_{k}/\partial\mathbf{s}_{k} & = \frac{1}{c}\begin{bmatrix}\cos(\theta_{\mathrm{Tx},k})+\cos(\theta_{\mathrm{Rx},k}),
\sin(\theta_{\mathrm{Tx},k})+\sin(\theta_{\mathrm{Rx},k})\end{bmatrix}^{\mathrm{T}},\:k> 0\\
\partial\theta_{\mathrm{Tx},k}/\partial\mathbf{s}_{k}& =\frac{1}{\Vert\mathbf{s}_{k}-\mathbf{q}\Vert_{2}}\begin{bmatrix}-\sin(\theta_{\mathrm{Tx},k}),\cos(\theta_{\mathrm{Tx},k})\end{bmatrix}^{\mathrm{T}},\:k> 0\\
\partial\theta_{\mathrm{Rx},k}/\partial\mathbf{p}& = \frac{1}{\Vert\mathbf{p}-\mathbf{s}_{k}\Vert_{2}}\begin{bmatrix}\sin(\pi-\theta_{\mathrm{Rx},k}), \cos(\pi-\theta_{\mathrm{Rx},k})\end{bmatrix}^{\mathrm{T}},\: k > 0\\
%\partial\theta_{\mathrm{Rx},r}/\partial\alpha & =-1,\:r\neq 0\\
\partial\theta_{\mathrm{Rx},k}/\partial\mathbf{s}_{k}& =-\frac{1}{\Vert\mathbf{p}-\mathbf{s}_{k}\Vert_{2}}\begin{bmatrix}\sin(\pi-\theta_{\mathrm{Rx},k}), \cos(\pi-\theta_{\mathrm{Rx},k})\end{bmatrix}^{\mathrm{T}},\:k> 0
\end{align*}
and $\partial\tilde{\mathbf{h}}^{\mathrm{T}}_{k}/\tilde{\mathbf{h}}_{k}=\mathbf{I}_{2}$ for $k \ge 0$. The rest of entries in $\mathbf{T}$ are zero. 

\subsection{Bounds on Position and Orientation Estimation Error}
The \ac{PEB} is obtained by inverting $\mathbf{J}_{\tilde{\boldsymbol{\eta}}}$, adding the diagonal entries of the $2\times 2$ sub-matrix, and taking the root square as:
\begin{equation}\label{PEBexpr}
\mathrm{PEB}=\sqrt{\mathrm{tr}\left\{[\mathbf{J}^{-1}_{\tilde{\boldsymbol{\eta}}}]_{1:2,1:2}\right\}},
\end{equation}
and the \ac{REB} is obtained as:
\begin{equation}\label{REBexpr}
\mathrm{REB}=\sqrt{[\mathbf{J}^{-1}_{\tilde{\boldsymbol{\eta}}}]_{3,3}},
\end{equation}
where the operations $[.]_{1:2,1:2}$ and $[.]_{3,3}$ denote the selection of the first $2\times 2$ sub-matrix and the third diagonal entry of $\mathbf{J}^{-1}_{\tilde{\boldsymbol{\eta}}}$, respectively. 
\subsection{The Effect of Multi-Path Components on Position and Orientation Estimation Error}
In this subsection, we discuss the effect of adding \ac{MPCs} for localization under different conditions.
%on the number of antenna elements in the BS and the MS, and on the system bandwidth. In principle, the received signal power is the same in every received antenna, and if the number of antennas in MS (i.e., $N_{r}$) increases, a greater gain can be obtained by combining the signals from all of them. This way one can use less powerful amplifier that leads to smaller gain per antenna element. Also, the received signal power in the direction $\theta_{\mathrm{Rx},r}$ using the steering vector from the direction $\theta_{\mathrm{Rx},s}$ with $r\neq s$ is very small, i.e., $\vert\mathbf{a}^{\mathrm{H}}_{\mathrm{Rx},n}(\theta_{\mathrm{Rx},r})\mathbf{a}_{\mathrm{Rx},n}(\theta_{\mathrm{Rx},s})\vert\ll 1$. 
As the number of antennas in the MS increases, the scalar product between steering vectors corresponding to different receive directions tends to vanish, i.e. $\vert\mathbf{a}^{\mathrm{H}}_{\mathrm{Rx},n}(\theta_{\mathrm{Rx},r})\mathbf{a}_{\mathrm{Rx},n}(\theta_{\mathrm{Rx},s})\vert\ll 1$ for $\theta_{\mathrm{Rx},r} \neq \theta_{\mathrm{Rx},s}$.
Also, increasing the number of antenna elements in the transmitter results in narrower beams and the spatial correlation between different beams is reduced. Moreover, as the system bandwidth % and eventually the number of subcarriers (i.e., $N$) 
increases, the different \ac{MPCs} coming from different scatterers can be more easily resolved. In other words, the MPCs can be considered to be orthogonal \cite{LeitingerJSAC2015,WitrisalSPM2016}. Consequently,  large $N_{t}$, $N_{r}$, and bandwidth lead to very small multipath cross-correlation terms in the \ac{FIM} \cite{DBLP:journals/corr/Abu-ShabanZASW17}.
%Consequently, the terms $\mathbf{A}_{\mathrm{Rx},n}(\theta_{\mathrm{Rx},r},\theta_{\mathrm{Rx},s})$, $A_{\mathbf{D}_{\mathrm{Rx,s}},n}(\theta_{\mathrm{Rx,r}},\theta_{\mathrm{Rx,s}})$, and $A_{\mathbf{D}_{\mathrm{Rx,r,s}},n}(\theta_{\mathrm{Rx,r}},\theta_{\mathrm{Rx,s}})$ in the FIM are very small. 
%Also, increasing the number of antenna elements in the transmitter results the narrower beams and the spatial correlation between different beams to be very small. This leads to very small values of $A^{(k)}_{\mathrm{Tx},\mathbf{F},n}(\tau_{r},\tau_{s},\theta_{\mathrm{Tx,s}},\theta_{\mathrm{Tx,r}})$, $A^{(l)}_{\mathbf{D}_{\mathrm{Tx},s},\mathbf{F},n}(\tau_{r},\tau_{s},\theta_{\mathrm{Tx,s}},\theta_{\mathrm{Tx,r}})$, $A^{(l)}_{\mathbf{D}_{\mathrm{Tx},r},\mathbf{F},n}(\tau_{r},\tau_{s},\theta_{\mathrm{Tx,s}},\theta_{\mathrm{Tx,r}})$, and $A_{\mathbf{Dd}_{\mathrm{Tx}},\mathbf{F},n}(\tau_{r},\tau_{s},\theta_{\mathrm{Tx,s}},\theta_{\mathrm{Tx,r}})$ for $r\neq s$ in the FIM. 
%Moreover, as the system bandwidth and eventually the number of subcarriers (i.e., $N$) increase, the effect of diffuse MPCs are reduced and different rays coming from different scatterers can be resolved. In other words, the MPCs are considered to be orthogonal \cite{LeitingerJSAC2015,WitrisalSPM2016}. 
Ignoring those terms, the approximate expression for the \ac{EFIM} of position and rotation angle $\mathbf{J}_{e}(\mathbf{p},\alpha)$ with large $N_{t}$, $N_{r}$, and bandwidth is\footnote{In computing \eqref{ans_com1}, we used the fact that the last two rows of $\mathbf{T}_{k,0}$ are zero for $k\neq 0$.} 
%\begin{equation}\label{ans_com1}
%\mathbf{J}_{e}(\mathbf{p},\alpha)\approx\sum_{k=0}^{K}\tilde{\mathbf{T}}_{k,0}\mathbf{\Lambda}_{e,k}\tilde{\mathbf{T}}^{\mathrm{T}}_{k,0}
%-\sum_{k=1}^{K}\tilde{\mathbf{T}}_{k,0}\mathbf{\Lambda}_{e,k}\tilde{\mathbf{T}}^{\mathrm{T}}_{k,k}
%\left(\tilde{\mathbf{T}}_{k,k}\mathbf{\Lambda}_{e,k}\tilde{\mathbf{T}}^{\mathrm{T}}_{k,k}\right)^{-1}\tilde{\mathbf{T}}_{k,k}\mathbf{\Lambda}_{e,k}\tilde{\mathbf{T}}^{\mathrm{T}}_{k,0},
%\end{equation}
%\begin{equation}\label{ans_com1}
%\mathbf{J}_{e}(\mathbf{p},\alpha)\approx\tilde{\mathbf{T}}_{0,0}\mathbf{\Lambda}_{e,0}\tilde{\mathbf{T}}^{\mathrm{T}}_{0,0}
%+\sum_{k=1}^{K}\tilde{\mathbf{T}}_{k,0}\mathbf{\Lambda}_{e,k}\tilde{\mathbf{T}}^{\mathrm{T}}_{k,0}-\tilde{\mathbf{T}}_{k,0}\mathbf{\Lambda}_{e,k}\tilde{\mathbf{T}}^{\mathrm{T}}_{k,k}
%\left(\tilde{\mathbf{T}}_{k,k}\mathbf{\Lambda}_{e,k}\tilde{\mathbf{T}}^{\mathrm{T}}_{k,k}\right)^{-1}\tilde{\mathbf{T}}_{k,k}\mathbf{\Lambda}_{e,k}\tilde{\mathbf{T}}^{\mathrm{T}}_{k,0},
%\end{equation}
\begin{equation}\label{ans_com1}
\mathbf{J}_{e}(\mathbf{p},\alpha)\approx\tilde{\mathbf{T}}_{0,0}\mathbf{\Lambda}_{e,0}\tilde{\mathbf{T}}^{\mathrm{T}}_{0,0}+
\sum_{k=1}^{K}\left[\mathbf{\Upsilon}_{e,k}\right]_{1:3,1:3},
\end{equation}
where
\begin{equation}\label{ans_com2}
\mathbf{\Upsilon}_{e,k}=\mathbf{T}_{k,0}\mathbf{\Psi}(\boldsymbol{\eta}_{k},\boldsymbol{\eta}_{k})\mathbf{T}^{\mathrm{T}}_{k,0}-\mathbf{T}_{k,0}\mathbf{\Psi}(\boldsymbol{\eta}_{k},\boldsymbol{\eta}_{k})\mathbf{T}^{\mathrm{T}}_{k,k}\left(\mathbf{T}_{k,k}\mathbf{\Psi}(\boldsymbol{\eta}_{k},\boldsymbol{\eta}_{k})\mathbf{T}^{\mathrm{T}}_{k,k}\right)^{-1}
\mathbf{T}_{k,k}\mathbf{\Psi}(\boldsymbol{\eta}_{k},\boldsymbol{\eta}_{k})\mathbf{T}^{\mathrm{T}}_{k,0},
\end{equation}
in which $\tilde{\mathbf{T}}_{0,0}$ is the $3\times 3$ sub-matrix in the transformation matrix $\mathbf{T}_{k,0}$ for $k=0$ in \eqref{TFIM6} containing the derivatives with respect to $\mathbf{p}$ and $\alpha$, $[.]_{1:3,1:3}$ denotes the selection of the first $3\times 3$ sub-matrix, and $\mathbf{\Lambda}_{e,0}$ denotes the EFIM of the delay, AOD, and AOA from LOS, i.e., $\{\tau_{0},\theta_{\mathrm{Tx},0},\theta_{\mathrm{Rx},0}\}$. From simulations, it is observed that the exact and approximate FIM lead to nearly identical PEBs, under the mentioned conditions. %In conclusion, for large $N_{t}$, $N_{r}$, and $N$ the paths become independent and the channel becomes more sparse. 
Hence, greedy techniques from compressed sensing, which extract path after path, are a natural tool for such scenarios.
%where $\tilde{\mathbf{T}}_{k,0}$ is the $3\times 3$ submatrix in the transformation matrix $\mathbf{T}_{k,0}$ in (26) with the derivatives with respect to $\mathbf{p}$ and $\alpha$, $\tilde{\mathbf{T}}_{k,k}$ is the $2\times 3$ submatrix in the transformation matrix $\mathbf{T}_{k,k}$ in (25) with the derivatives with respect to $\mathbf{s}_{k}$, and $\mathbf{\Lambda}_{e,k}$ denotes the EFIM of the delay, AOD, and AOA from the $k$-th path, i.e., $\{\tau_{k},\theta_{\mathrm{Tx},k},\theta_{\mathrm{Rx},k}\}$. 
In the LOS case, \eqref{ans_com1} only contains the term corresponding to $k=0$, i.e., the first term. When MPCs are present, the terms corresponding to $k\geq 1$ appear, i.e., the second summand in \eqref{ans_com1}, which contains terms that are added and others that are subtracted (because the scatterer location is an additional parameter that has to be estimated for each MPC \cite[eq. (3.59)]{LeitingerPhD2016}). The additive terms imply that the presence of MPCs help in the estimation of the MS localization, as they add information to the EFIM. In general the contribution of the MPCs results in a positive contribution to the FIM, and hence in a reduction of the CRB as shown in papers \cite{LeitingerJSAC2015,WitrisalSPM2016}. It is only in the cases where the MPCs heavily overlap, specially with the LOS, in the directional and time domains that the negative terms are dominant, and then the presence of MPCs degrades the MS localization. 
%Fig. \ref{Evolution_vs_K} shows the evolution of the PEB by adding the scatterers based on the approximate and exact expressions of the FIM. It is clear the PEB is reduced by adding the scatterers. Moreover, the PEB obtained using the approximate expression of the FIM follows the PEB based on the exact expression of the FIM with very small deviation shown for $K=1$.
%It is only in the cases where the MPCs heavily overlap, specially with the LOS, in the directional or time domain that the negative terms are dominant, and then the presence of MPCs degrades the MS localization.
%In \eqref{ans_com1}, the first term represents information obtained by the LOS and the second term adds the effect of MPCs. Since the channel paths and the scatterers still need to be estimated, their effects are subtracted from a positive value in the second term in \eqref{ans_com1}. The combinations of these effects increases $\mathbf{J}_{e}(\mathbf{p},\alpha)$ that results reducing the PEB. In conclusion, for large $N_{t}$, $N_{r}$, and $N$ the paths become independent and the channel becomes more sparse. Hence, greedy techniques from compressed sensing are a natural tool for such scenarios. For the case that the beams coming from different scatterers are too close and the beams cannot be resolved or the case that $N_{t}$, $N_{r}$, and $N$ are not sufficiently large, adding the MPCs theoretically degrades the performance. However, in practice one can consider close beams as one beam and estimate the corresponding parameters for one beam that solves the degradation due to adding non-resolvable MPCs. 
\section{Position and Orientation Estimation: Estimator in Beamspace}

Next, we propose the use of a beamspace channel transformation in order to estimate the channel parameters in \eqref{Receivedb1}. The considered beamspace representation of the channel reduces the complexity by exploiting the sparsity of the \ac{mm-wave} MIMO channel. If the fractional bandwidth and the number of antennas are not violating the condition for the small array dispersion \cite{widebandbrady}, there exists a common sparse support across all subcarriers. Consequently, the \ac{DCS-SOMP} method  from \cite{Duarte2} can be applied for the estimation of \ac{AOA}, \ac{AOD}, and \ac{TOA}. As the estimates of \ac{AOA} and \ac{AOD} are limited to lie on a grid defined by the transformation, we apply a refinement of the estimates of all parameters using the \ac{SAGE} algorithm. Finally, we invoke the \ac{EXIP} to solve for the position $\mathbf{p}$ and orientation $\alpha$.

\subsection{Beamspace Channel Representation}

We introduce the $N_t \times N_t$ transformation matrix, uniformly sampling the  virtual spatial angles \cite{BspaceSayeedx}
\begin{align*}
\mathbf{U}_{\mathrm{Tx}}&\triangleq\left[\mathbf{u}_{\mathrm{Tx}}({-(N_{t}-1)/2}),\ldots,\mathbf{u}_{\mathrm{Tx}}({(N_{t}-1)/2})\right],\\
\mathbf{u}_{\mathrm{Tx}}(p)&\triangleq \begin{bmatrix}
e^{-j2\pi \frac{N_{t}-1}{2} \frac{p}{{N_{t}}}},\ldots,e^{j2\pi \frac{N_{t}-1}{2} \frac{p}{N_{t}}}
\end{bmatrix}^{\mathrm{T}},
\end{align*}
where we assumed $N_t$ to be even. 
Similarly, we define the $N_r \times N_r$ matrix $\mathbf{U}_{\mathrm{Rx}}$. Both $\mathbf{U}_{\mathrm{Tx}}$ and $\mathbf{U}_{\mathrm{Rx}}$ are unitary matrices.  The partial virtual representation of the channel with respect to the angular domain can be written as 
\begin{align}
\check{\mathbf{H}}[n]&=\mathbf{U}_{\mathrm{Rx}}^{\mathrm{H}}\mathbf{H}[n]\mathbf{U}_{\mathrm{Tx}}\label{BWTransceiver1a}\\
& = \sum_{k=0}^{K}\gamma_{n}(h_{k},\tau_{k})\mathbf{U}_{\mathrm{Rx}}^{\mathrm{H}}\mathbf{a}_{\mathrm{Rx},n}(\theta_{\mathrm{Rx},k})\mathbf{a}^{\mathrm{H}}_{\mathrm{Tx},n}(\theta_{\mathrm{Tx},k})\mathbf{U}_{\mathrm{Tx}}.
%\check{H}_{i,m}[n]&=\sum_{k=0}^{K}\gamma_{n}(h_{k},\tau_{k})\varsigma_{N_{r}}(\phi_{\mathrm{Rx},k,n}-\frac{i}{N_{r}})\varsigma_{N_{t}}(\phi_{\mathrm{Tx},k,n}-\frac{m}{N_{t}}),\\
\end{align}
It is readily verified that \cite{widebandbrady} 
\begin{align}
[\check{\mathbf{H}}[n]]_{i,i'}&=\sum_{k=0}^{K}\gamma_{n}(h_{k},\tau_{k})\chi_{r}\big(\frac{d}{\lambda_{n}}\sin(\theta_{\mathrm{Rx},k})-\frac{i}{N_{r}}\big)\chi_{{t}}\big(\frac{d}{\lambda_{n}}\sin(\theta_{\mathrm{Tx},k})-\frac{i'}{N_{t}}\big),\label{BWTransceiver1b}
\end{align}
for $-(N_r-1)/2 \le i \le (N_r-1)/2$ and $-(N_t-1)/2 \le i' \le (N_t-1)/2$. We have introduced
\begin{align}
\chi_t(\phi) & = \frac{\sin(\pi N_{t}\phi)}{\sqrt{N_{t}}\sin(\pi\phi)},\\
\chi_r(\phi) & = \frac{\sin(\pi N_{r}\phi)}{\sqrt{N_{r}}\sin(\pi\phi)}.
\end{align}
From \eqref{BWTransceiver1b},  it is observed that $\check{\mathbf{H}}[n]$ is approximately sparse, since `strong' components are only present in the directions of $\{\theta_{\mathrm{Tx},k}\}$ and $\{\theta_{\mathrm{Rx},k}\}$. 

Stacking the observation $\mathbf{y}^{(g)}[n]$ from \eqref{Receivedb1}, we obtain 
\begin{equation}\label{BWTransceiver2x}
\check{\mathbf{y}}[n]=\mathbf{\Omega}[n]\check{\mathbf{h}}[n]+\check{\mathbf{n}}[n],
\end{equation}
where
\begin{align}
\mathbf{\Omega}[n]&=\begin{bmatrix}
\mathbf{\Omega}^{(1)}[n]\\\vdots\\\mathbf{\Omega}^{(G)}[n]
\end{bmatrix},\\
\mathbf{\Omega}^{(g)}[n]&=(\mathbf{Z}^{(g)}_{\mathrm{Tx}}[n])^{\mathrm{T}}\otimes\mathbf{U}_{\mathrm{Rx}},\\
\mathbf{Z}^{(g)}_{\mathrm{Tx}}[n]&=\mathbf{U}^{\mathrm{H}}_{\mathrm{Tx}}\mathbf{F}^{(g)}[n]\mathbf{x}^{(g)}[n],\\
\check{\mathbf{h}}[n] &= \mathrm{vec}(\check{\mathbf{H}}[n]).
\end{align}
Hence, since $\check{\mathbf{h}}[n]$ is an approximately sparse vector, we can interpret solving \eqref{BWTransceiver2x} for $\check{\mathbf{h}}[n]$ as a {CS} problem, allowing us to utilize tools from that domain. In principle, the columns of $\mathbf{U}_{\mathrm{Tx}}$ and $\mathbf{U}_{\mathrm{Rx}}$ corresponding to non-zero entries of the sparse vector $\check{\mathbf{h}}[n]$ correspond to coarse estimates of the AOA/AOD, while the entries in $\check{\mathbf{h}}[n]$ are estimates of $\gamma_{n}(h_{k},\tau_{k})$ (including the effect of the functions $\chi_t(\cdot)$ and $\chi_r(\cdot)$). The latter values can then be used to estimate $\tau_{k}$ for each path.  Since the  vectors $\check{\mathbf{h}}[n]\in\mathbb{C}^{N_{r}N_{t}\times 1}$, for $i=1,\ldots,N$, corresponding to the sensing matrix $\mathbf{\Omega}[n]$ in \eqref{BWTransceiver2x} are approximately jointly $(K+1)$-sparse, i.e., the support of $\check{\mathbf{h}}[n]$ does not vary significantly from subcarrier to subcarrier, we can use specialized techniques, such as \ac{DCS-SOMP} for estimating all $\check{\mathbf{h}}[n]$ jointly in an efficient manner. 

Based on the above discussion, we propose to use the following approach:
\begin{enumerate}
\item Coarse estimation of AOA/AOD using a modified \ac{DCS-SOMP} algorithm. %Since the estimated value of the AOA/AOD are limited to the grid defined by $\mathbf{U}_{\mathrm{Tx}}$ and $\mathbf{U}_{\mathrm{Rx}}$, these estimates are considered as coarse. 
\item Fine estimation using the SAGE algorithm, initialized by the coarse estimates.
\item Estimation of the position and orientation. 
\end{enumerate}

\subsubsection*{Remark}
The above sparse representation is not unique. Another representation could rely on a sparse vector of length $N_{t}\times N_{r}\times N$, where each entry would then correspond to an AOA/AOD/TOA triplet. However, the complexity of such an approach would be significantly higher, since $N$ is generally a large number. 

\subsection{Step 1: Coarse Estimation of Channel Parameters using DCS-SOMP}\label{EstRef}
The first stage of the algorithm involves calling the DCS-SOMP algorithm, providing estimates of the number of paths, the AOA/AOD, and estimates of $\check{\mathbf{h}}[n]$. For the sake of completeness, the steps of DCS-SOMP can be found in Algorithm \ref{algor0_det}. We note that the algorithm is rank-blind as it does not assume knowledge of the number of the paths (i.e., $K+1$) \cite{Davies}. Since $K+1$ is unknown, we use the change of residual fitting error $\sum_{n=0}^{N-1}\Vert\mathbf{r}_{t-1}[n]-\mathbf{r}_{t-2}[n]\Vert^{2}_{2}$ at each iteration $t$ to a threshold $\delta$. The value for $\delta$ is obtained using a similar procedure as in \cite{Marzi}:
\begin{equation}\label{reqproof3}
\delta=N_0\gamma^{-1}\left(N,\Gamma(N)(1-{P}_{\mathrm{fa}})^{{1}/({N_{r}N_{t}})}\right),
\end{equation}
in which $\gamma^{-1}\left(N,x\right)$ denotes the inverse of the incomplete gamma distribution, $\Gamma(N)$ is the gamma function, and ${P}_{\mathrm{fa}}$ is the false alarm probability. 

%based on the noise power, to determine the terminationof the algorithm. In particular, once all signal components have been detected, only noise remains with a known power, allowing us to choose a value of $\delta$ for a certain probability of false alarm. 

\begin{algorithm}[h!]
\caption{Modified DCS-SOMP\label{algor0_det}}
\textbf{Input:} Recieved signals $\check{\mathbf{y}}[n]$, sensing matrix $\mathbf{\Omega}[n]$, and the threshold $\delta$.\\
\textbf{Output:}  estimates of $K$, ${\theta}_{\mathrm{Tx},k}$, ${\theta}_{\mathrm{Rx},k}$, $\check{\mathbf{h}}[n]$, $n=0,\ldots,N-1.$
\begin{algorithmic}[1]
\STATE For $n=0,\ldots,N-1$, the residual vectors are set to $\mathbf{r}_{-1}[n]=\mathbf{0}$ and $\mathbf{r}_{0}[n]=\check{\mathbf{y}}[n]$, the
orthogonalized coefficient vector $\hat{\boldsymbol{\beta}}_{n}=\mathbf{0}$, $\mathcal{K}_{0}$ is chosen to be an empty set, and iteration index $t=1$. $\boldsymbol{\omega}_{m}[n]$ is the $m$-th column of measurement matrix $\mathbf{\Omega}[n]$. 
\WHILE{ $\sum_{n=0}^{N-1}\Vert\mathbf{r}_{t-1}[n]-\mathbf{r}_{t-2}[n]\Vert^{2}_{2}>\delta$} 
\STATE Find AOA/AOD pair
\begin{align}
\tilde{n}_{t}& =\underset{m=1,\ldots,N_{r}N_{t}}
{\mathrm{argmax}} \:\sum_{n=0}^{N-1}\frac{\vert 
\boldsymbol{\omega}_{m}^{\mathrm{H}}[n]\mathbf{r}_{t-1}[n]\vert}{\Vert\boldsymbol{\omega}_{m}[n]\Vert_{2}},\label{tinex1ee}\\
n_{\mathrm{Tx},t}&=\lceil \tilde{n}_{t}/N_{r}\rceil, ~~
n_{\mathrm{Rx},t}=\mathrm{mod}(\tilde{n}_{t}-1,N_{r})+1,\\
\hat{\theta}^{(0)}_{\mathrm{Tx},t} & =\arcsin\left((\lambda_{c}/d)(n_{\mathrm{Tx},t}-{(N_{t}-1)/2}-1)/N_{t}\right),\label{tinex1}\\
 \hat{\theta}^{(0)}_{\mathrm{Rx},t}& =\arcsin\left((\lambda_{c}/d)(n_{\mathrm{Rx},t}-{(N_{r}-1)/2}-1)/N_{r}\right).\label{rinex1}
\end{align}
\STATE Update AOA/AOD set of indices $\mathcal{K}_{t}=\mathcal{K}_{t-1}\cup\{\tilde{n}\}$.
\STATE Orthogonalize the selected basis vector:
\begin{equation}\label{BWTransceiver2zfh}
\boldsymbol{\rho}_{t}[n]=\boldsymbol{\omega}_{\tilde{n}_{t}}[n]-\sum_{\tilde{t}=0}^{t-1}
\frac{\boldsymbol{\omega}^{\mathrm{H}}_{\tilde{n}_{t}}[n]\boldsymbol{\rho}_{\tilde{t}}[n]}{\Vert\boldsymbol{\rho}_{\tilde{t}}[n]\Vert_{2}}\boldsymbol{\rho}_{\tilde{t}}[n].
\end{equation}
\STATE Update the residual vector $\mathbf{r}_{t}[n]$  by subtracting the effect of chosen columns from $\mathbf{r}_{t-1}[n]$: $\mathbf{r}_{t}[n]=\mathbf{r}_{t-1}[n]-\hat{\beta}_{n}(t)\boldsymbol{\rho}_{t}[n]$, 
%\begin{equation}\label{BWTransceiver2zdy}
%\mathbf{r}_{t}[n]=\mathbf{r}_{t-1}[n]-\hat{\beta}_{n}(t)\boldsymbol{\rho}_{t}[n],
%\end{equation}
where
\begin{equation}\label{BWTransceiver2zdyb}
\hat{\beta}_{n}(t)=\frac{\boldsymbol{\rho}^{\mathrm{H}}_{t}[n]\mathbf{r}_{t-1}[n]}{\Vert\boldsymbol{\rho}_{t}[n]\Vert^{2}_{2}}.
\end{equation}
\STATE $t=t+1$.
\ENDWHILE
%\STATE Find $\mathbf{h}_{\mathrm{v},t}[n]$ for $n=0,\ldots,N-1$ based on \eqref{BWTransceiver2z}.
\STATE Perform QR factorization of the mutilated basis $\mathbf{\Omega}_{\mathcal{K}_{t}}[n]=[\boldsymbol{\omega}_{\tilde{n}_{1}}[n],\ldots,\boldsymbol{\omega}_{\tilde{n}_{\hat{K}+1}}[n]]=\mathbf{\Upsilon}[n]\mathbf{R}[n]$ where $\mathbf{\Upsilon}[n]=[\boldsymbol{\rho}_{1}[n],\ldots,\boldsymbol{\rho}_{\hat{K}+1}[n]]$ and $\mathbf{R}[n]$ is an upper triangular matrix. Since $\mathbf{\Omega}_{\mathcal{K}_{t}}[n]\hat{\check{\mathbf{h}}}[n]=\mathbf{\Upsilon}[n]\mathbf{R}[n]\hat{\check{\mathbf{h}}}[n]=\mathbf{\Upsilon}[n]\hat{\boldsymbol{\beta}}_{n}$, we obtain
\begin{equation}\label{BWTransceiver2z}
\hat{\check{\mathbf{h}}}[n]=\mathbf{R}^{-1}[n]\hat{\boldsymbol{\beta}}_{n}.
\end{equation}
%\STATE Find gains using \eqref{BWTransceiver2tzsfzxe2}, and TOA using \eqref{BWTransceiver2tzsfzxe3}.
\end{algorithmic}
\end{algorithm}

For each path $k = 0,\ldots, \hat{K}$, we can now write 
\begin{equation}\label{BWTransceiver2tz0}
\hat{\check{\mathbf{h}}}^{(k)}=\tilde{h}_{k}\mathbf{A}(\tau_{k})\mathbf{z}^{(k)}+\mathbf{v}^{(k)},
\end{equation}
where  $\hat{\check{\mathbf{h}}}^{(k)}=[\hat{\check{h}}^{(k)}[0],\ldots,\hat{\check{h}}^{(k)}[N-1]]^{\mathrm{T}}$ in which $\hat{\check{h}}^{(k)}[n]$ is the entry on subcarrier $n$, related to the $k$-th path found in Algorithm \ref{algor0_det}, $\mathbf{A}(\tau_{k})=\mathrm{diag}\{1,\ldots,e^{-j2\pi (N-1)\tau_{k}/(NT_{s})}\}$, $\mathbf{v}_{k}$ is the $N\times 1$ noise vector, and $\mathbf{z}^{(k)}$ has entries
\begin{equation}\label{Refine2}
z_{n}(k)\triangleq\mathbf{u}^{\mathrm{H}}_{\mathrm{Rx}}(\frac{n_{\mathrm{Rx},k}-{(N_{r}-1)/2}-1}{N_{r}})\mathbf{a}_{\mathrm{Rx},n}(\hat{\theta}^{(0)}_{\mathrm{Rx},k})\mathbf{a}^{\mathrm{H}}_{\mathrm{Tx},n}(\hat{\theta}^{(0)}_{\mathrm{Tx},k})\mathbf{u}_{\mathrm{Tx}}(\frac{n_{\mathrm{Tx},k}-{(N_{t}-1)/2}-1}{N_{t}}).
\end{equation}
For the purpose of coarse estimation, we ignore the dependence on $n$ in \eqref{Refine2}, leading to the simple model 
\begin{align}
\hat{\check{\mathbf{h}}}^{(k)}=\tilde{h}_{k}{z}^{(k)}\mathbf{a}(\tau_k)+\mathbf{v}^{(k)},
\end{align}
 where $\mathbf{a}(\tau_k) = [1,\ldots,e^{-j2\pi (N-1)\tau_{k}/(NT_{s})}]^{\mathrm{T}}$ and ${z}^{(k)}$ is as in \eqref{Refine2}, but considering only $\lambda_c$  instead of $\lambda_n$. From this model, we can recover $\tau_{k}$ and $\tilde{h}_{k}$ by solving a \ac{LS}  problem
\begin{equation}\label{BWTransceiver2tzsf}
[\hat{\tau}^{(0)}_{k},\hat{\tilde{h}}^{(0)}_{k}]=\underset{\tau_{k},\tilde{h}_{k}}{\mathrm{argmin}}\:\:\Vert\hat{\check{\mathbf{h}}}^{(k)}-\tilde{h}_{k}{z}^{(k)}\mathbf{a}(\tau_k)\Vert_{2}^{2}.
\end{equation}
Solving for ${\tilde{h}}_{k}$ yields
\begin{align}\label{BWTransceiver2tzsfzxe2}
\hat{\tilde{h}}^{(0)}_{k}= \frac{\mathbf{a}^{\mathrm{H}}(\tau_k)\hat{\check{\mathbf{h}}}^{(k)}}{{z}^{(k)}N}.
\end{align}
Substituting \eqref{BWTransceiver2tzsfzxe2} into \eqref{BWTransceiver2tzsf} and expanding the square allows us to solve for ${\tau}_{k}$: 
\begin{align}\label{BWTransceiver2tzsfzxe3}
\hat{\tau}^{(0)}_{k}=\underset{\tau_{k}}{\mathrm{argmax}}\:\: |\mathbf{a}^{\mathrm{H}}(\tau_k)\hat{\check{\mathbf{h}}}^{(k)}|^2.
\end{align}

\subsection{Step 2: Fine Estimation of Channel Parameters using SAGE}
Channel parameter estimates are refined in an iterative procedure, which is initialized by the estimates from step 1. In principle, we can perform an iterative ascent algorithm directly on the log-likelihood function associated with the model \eqref{BWTransceiver2x}. However, this requires a multi-dimensional minimization and computationally complex solutions. A more practical approach is to use the \ac{SAGE} algorithm with the incomplete data space in \eqref{BWTransceiver2x} as the superposition of $K+1$ complete data space $\check{\mathbf{y}}_{k}[n]$ as:
\begin{equation}\label{Refine1}
\check{\mathbf{y}}[n]=\sum_{k=0}^{\hat{K}}\underbrace{\mathbf{\Omega}[n]\check{\mathbf{h}}_{k}[n]+\check{\mathbf{n}}_{k}[n]}_{\check{\mathbf{y}}_{k}[n]},
\end{equation}
where $\check{\mathbf{h}}_{k}[n]$ denotes the vectorized form of $\check{\mathbf{H}}_{k}[n]=\mathbf{U}_{\mathrm{Rx}}^{\mathrm{H}}\mathbf{H}_{k}[n]\mathbf{U}_{\mathrm{Tx}}$ with $\mathbf{H}_{k}[n]$ being the corresponding term for the $k$-th path in the channel frequency response $\mathbf{H}[n]$ in \eqref{Channel1}. Writing \eqref{Refine1} for all the subcarriers results in:
\begin{equation}\label{Refine1x}
\check{\mathbf{y}}=\sum_{k=0}^{\hat{K}}\underbrace{\check{\mathbf{\Omega}}\check{\mathbf{h}}_{k}
+\check{\mathbf{n}}_{k}}_{\check{\mathbf{y}}_{k}},
\end{equation} 
where 
%$\bar{\mathbf{\Omega}}=\mathrm{diag}\left\{\mathbf{\Omega}[0],\ldots,\mathbf{\Omega}[N-1]\right\}$, $\bar{\mathbf{y}}=\left[\bar{\mathbf{y}}^{\mathrm{T}}[0],\ldots,\bar{\mathbf{y}}^{\mathrm{T}}[N-1]\right]^{\mathrm{T}}$, $\bar{\mathbf{h}}_{\mathrm{v},k}=\left[\bar{\mathbf{h}}^{\mathrm{T}}_{\mathrm{v},k}[0],\ldots,\bar{\mathbf{h}}^{\mathrm{T}}_{\mathrm{v},k}[N-1]\right]^{\mathrm{T}}$, and $\bar{\mathbf{n}}_{k}=\left[\bar{\mathbf{n}}_{k}^{\mathrm{T}}[0],\ldots,\bar{\mathbf{n}}_{k}^{\mathrm{T}}[N-1]\right]^{\mathrm{T}}$.
\begin{align*}
\check{\mathbf{\Omega}}&=\mathrm{diag}\left\{\mathbf{\Omega}[0],\ldots,\mathbf{\Omega}[N-1]\right\},\\
\check{\mathbf{y}}&=\left[\check{\mathbf{y}}^{\mathrm{T}}[0],\ldots,\check{\mathbf{y}}^{\mathrm{T}}[N-1]\right]^{\mathrm{T}},\\
\check{\mathbf{h}}_{k}&=\left[\check{\mathbf{h}}^{\mathrm{T}}_{k}[0],\ldots,\check{\mathbf{h}}^{\mathrm{T}}_{k}[N-1]\right]^{\mathrm{T}},\\
\check{\mathbf{n}}_{k}&=\left[\check{\mathbf{n}}_{k}^{\mathrm{T}}[0],\ldots,\check{\mathbf{n}}_{k}^{\mathrm{T}}[N-1]\right]^{\mathrm{T}}.
\end{align*}
In the $(m+1)$-th iteration where $m$ is the iteration index, the expectation and maximization steps are performed as described below. For the initialization of the iterative procedure, we use the AOA/AOD, TOA, and channel coefficients from the detection phase using $\hat{\theta}^{(0)}_{\mathrm{Tx},k}$ and $\hat{\theta}^{(0)}_{\mathrm{Rx},k}$ obtained from \eqref{tinex1} and \eqref{rinex1}, respectively, $\hat{\tau}^{(0)}_{k}$ computed from \eqref{BWTransceiver2tzsfzxe3}, and the corresponding coefficient obtained from \eqref{BWTransceiver2tzsfzxe2}. 

\subsubsection*{Expectation step} We compute the conditional expectation of the hidden data space $\check{\mathbf{y}}_{k}$ log-likelihood function based on the previous estimation $\hat{\boldsymbol{\eta}}^{(m)}$ and the incomplete data space $\check{\mathbf{y}}$ as:
\begin{equation}\label{ExpectationS1}
Q(\boldsymbol{\eta}_{k}\vert\hat{\boldsymbol{\eta}}^{(m)})\triangleq \mathbb{E}\left[\ln f(\check{\mathbf{y}}_{k}\vert\boldsymbol{\eta}_{k},\{\hat{\boldsymbol{\eta}}^{(m)}_{l}\}_{l\neq k})\vert\check{\mathbf{y}},\hat{\boldsymbol{\eta}}^{(m)}\right].
\end{equation}
For $k=0,\ldots,\hat{K}$, we obtain 
\begin{equation}\label{ExpectationS2}
Q(\boldsymbol{\eta}_{k}\vert\hat{\boldsymbol{\eta}}^{(m)})\propto -\Vert\hat{\mathbf{z}}^{(m)}_{k}-\check{\boldsymbol{\mu}}(\boldsymbol{\eta}_{k})\Vert_{2}^{2},
\end{equation}
where $\check{\boldsymbol{\mu}}(\boldsymbol{\eta}_{k})=\check{\mathbf{\Omega}}\check{\mathbf{h}}_{k}$, and
\begin{equation}\label{ExpectationS3}
\hat{\mathbf{z}}^{(m)}_{k}=\check{\mathbf{y}}-\sum_{l\neq k, l=0}^{\hat{K}}\check{\boldsymbol{\mu}}(\hat{\boldsymbol{\eta}}^{(m)}_{l}).
\end{equation}

\subsubsection*{Maximization step} The goal is to find $\boldsymbol{\eta}_{k}$ such that \eqref{ExpectationS2} is maximized. In other words, we have
\begin{equation}\label{MaximizationS1}
\hat{\boldsymbol{\eta}}^{(m+1)}_{k}=\underset{\boldsymbol{\eta}_{k}}{\mathrm{argmax}}\:\:Q(\boldsymbol{\eta}_{k}\vert\hat{\boldsymbol{\eta}}^{(m)}).
\end{equation}
Solving \eqref{MaximizationS1} directly for $\boldsymbol{\eta}_{k}$ is analytically complex due to the fact that it is hard to compute the gradient and Hessian with respect to $\boldsymbol{\eta}_{k}$. Instead, we update the parameters $\hat{\theta}^{(m+1)}_{\mathrm{Tx},k}$, $\hat{\theta}^{(m+1)}_{\mathrm{Rx},k}$, $\hat{\tau}^{(m+1)}_{k}$, and $\hat{\tilde{h}}^{(m+1)}_{k}$ sequentially using Gauss-Seidel-type iterations \cite{Ortega}.

\subsection{Step 3: Conversion to Position and Rotation Angle Estimates}
As a final step, based on the refined estimates of AOA/AOD/TOA from step 2, here we show how the position and orientation of the MS is recovered. Four scenarios are considered: LOS, NLOS, OLOS, and unknown condition. 
\begin{itemize}
\item \textbf{LOS:}  When $\hat{K}=1$ and we are in LOS condition, the expressions \eqref{TFIM2xy1}, \eqref{TFIM2xy2}, and \eqref{TFIM2xy5} describe a mapping $\boldsymbol{\eta} = \boldsymbol{f}_{\mathrm{los}}(\tilde{\boldsymbol{\eta}})$. The classical invariance principle of estimation theory is invoked to prove the equivalence of minimizing the \ac{ML} criterion in terms of either $\boldsymbol{\eta}_{0}$ or $\tilde{\boldsymbol{\eta}}_{0}$ \cite{Zacks}. Consequently, the estimated values of $\hat{\mathbf{p}}$ and $\hat{\alpha}$ are obtained directly from   
\begin{align}\label{EXIP2}
\hat{\mathbf{p}}&=\mathbf{q}+c\hat{\tau}_{0}[\cos(\hat{\theta}_{\mathrm{Tx},0}),\sin(\hat{\theta}_{\mathrm{Tx},0})]^{\mathrm{T}},\\
\hat{\alpha}&=\pi+\hat{\theta}_{\mathrm{Tx},0}-\hat{\theta}_{\mathrm{Rx},0}.
\end{align}
%and $\hat{\alpha}=\pi+\hat{\theta}_{\mathrm{Tx},0}-\hat{\theta}_{\mathrm{Rx},0}$ where $\hat{d}_{0}=c\hat{\tau}_{0}$ denotes the distance between the MS and the BS in the LOS.

\item \textbf{NLOS:} For the case with $\hat{K}$ scatterers and a \ac{LOS} path, the \ac{EXIP} can be used, as \eqref{TFIM2xy1}--\eqref{TFIM2xy5} describe a mapping $\boldsymbol{\eta} = \boldsymbol{f}_{\mathrm{nlos}}(\tilde{\boldsymbol{\eta}})$. Consequently, the estimated $\hat{\tilde{\boldsymbol{\eta}}}$ obtained as
\begin{equation}\label{EXIP4}
\hat{\tilde{\boldsymbol{\eta}}}=\underset{\tilde{\boldsymbol{\eta}}}{\mathrm{argmin}}\:
\underbrace{\left(\hat{\boldsymbol{\eta}}-\boldsymbol{f}_{\mathrm{nlos}}(\tilde{\boldsymbol{\eta}})\right)^{\mathrm{T}}\mathbf{J}_{\hat{\boldsymbol{\eta}}}\left(\hat{\boldsymbol{\eta}}-\boldsymbol{f}_{\mathrm{nlos}}(\tilde{\boldsymbol{\eta}})\right)}_{v_{\mathrm{nlos}}(\tilde{\boldsymbol{\eta}})},
\end{equation}
is asymptotically (w.r.t.~$G\times N$) equivalent to the ML estimate of the transformed parameter $\tilde{\boldsymbol{\eta}}$ \cite{Stoicapp,Swindlehurstt}. Note that $\mathbf{J}_{\boldsymbol{\eta}}$ could be replaced by the identity matrix, leading also to a meaningful estimator of $\tilde{\boldsymbol{\eta}}$, although with probably slightly larger \ac{RMSE}. 
 The \ac{LMA} can be used to solve \eqref{EXIP4} \cite{Levenberg,Marquardt}, initialized as follows: 
we first estimate $\hat{\mathbf{p}}$ and $\hat{\alpha}$  from the \ac{LOS} path (i.e., the path with the smallest delay). Then, for the first-order reflection $\hat{\mathbf{s}}_{k}$ 
can be obtained by the intersection of the following two lines: $\tan(\pi-(\hat{\theta}_{\mathrm{Rx},k}+\hat{\alpha}))=(\hat{p}_{y}-s_{1,y})/(\hat{p}_{x}-s_{1,x})$ and $\tan(\hat{\theta}_{\mathrm{Tx},k})=(s_{1,y}-q_{y})/(s_{1,x}-q_{x})$.

\item \textbf{OLOS:} For the case with $\hat{K}$ scatterers and no LOS path, the \ac{EXIP} could be used, as \eqref{TFIM2xy1b}, \eqref{TFIM2xy3}, and \eqref{TFIM2xy4}  describe a mapping $\boldsymbol{\eta}_{\mathrm{olos}} = \boldsymbol{f}_{\mathrm{olos}}(\tilde{\boldsymbol{\eta}}_{\mathrm{olos}})$. 
Consequently, the estimated $\hat{\tilde{\boldsymbol{\eta}}}_{\mathrm{olos}}$ obtained as
\begin{equation}\label{EXIP4b}
\hat{\tilde{\boldsymbol{\eta}}}_{\mathrm{olos}}=\underset{\tilde{\boldsymbol{\eta}}_{\mathrm{olos}}}{\mathrm{argmin}}\:
\underbrace{\left(\hat{\boldsymbol{\eta}}_{\mathrm{olos}}-\boldsymbol{f}_{\mathrm{olos}}(\tilde{\boldsymbol{\eta}}_{\mathrm{olos}})\right)^{\mathrm{T}}\mathbf{J}_{\hat{\boldsymbol{\eta}}_{\mathrm{olos}}}\left(\hat{\boldsymbol{\eta}}_{\mathrm{olos}}-\boldsymbol{f}_{\mathrm{olos}}(\tilde{\boldsymbol{\eta}}_{\mathrm{olos}})\right)}_{v_{\mathrm{olos}}(\tilde{\boldsymbol{\eta}}_{\mathrm{olos}})},
\end{equation}
is asymptotically equivalent to the ML estimate of the transformed parameter $\tilde{\boldsymbol{\eta}}_{\mathrm{olos}}$ where $\mathbf{J}_{\hat{\boldsymbol{\eta}}_{\mathrm{olos}}}$ denotes the \ac{FIM} of $\boldsymbol{\eta}_{\mathrm{olos}}$. The estimated parameters from the \ac{NLOS} links could be used to initialize $\tilde{\boldsymbol{\eta}}_{\mathrm{olos}}$ for the application of the \ac{LMA} algorithm. The process is slightly more involved than under NLOS. We consider different trial values of $\alpha$, with a resolution $\Delta \alpha$ over a range $[-\alpha_m,+\alpha_m]$ of possible rotation values. For each trial value $\hat{\alpha}_{\mathrm{trial}}$, we can find a corresponding estimate of $\mathbf{p}$. For instance, by solving a set of linear equations for two paths:
\begin{equation}\label{EXIP5b}
\mathbf{p}=\mathbf{q}+d_{k,1}\begin{bmatrix}
\cos(\hat{\theta}_{\mathrm{Tx},k})\\\sin(\hat{\theta}_{\mathrm{Tx},k})
\end{bmatrix}+(c\hat{\tau}_{k}-d_{k,1})\begin{bmatrix}
\cos(\hat{\theta}_{\mathrm{Rx},k}+\hat{\alpha}_{\mathrm{trial}})\\-\sin(\hat{\theta}_{\mathrm{Rx},k}+\hat{\alpha}_{\mathrm{trial}})
\end{bmatrix},\: k \in \{ k_1,k_2 \}
\end{equation} 
where $d_{k,1}$ was introduced in Fig.~\ref{NLOS_Link}. After solving \eqref{EXIP5b} for $[\mathbf{p},d_{1,1},d_{2,1}]$, it is straightforward to determine the scatterer locations (as was done in the NLOS case). For each trial value $\hat{\alpha}_{\mathrm{trial}}$, we can then apply the \ac{LMA}  to \eqref{EXIP4b} to obtain $\hat{\tilde{\boldsymbol{\eta}}}_{\mathrm{olos}}$. The solution $\hat{\tilde{\boldsymbol{\eta}}}_{\mathrm{olos}}$ with the smallest $v_{\mathrm{olos}}(\tilde{\boldsymbol{\eta}}_{\mathrm{olos}})$ (with respect to all possible trial value  $\hat{\alpha}_{\mathrm{trial}}$) is then retained. Clearly, there is a performance/complexity trade-off based on the choice of $\Delta \alpha$.   It is readily seen that to obtain estimates of all parameters, at least three scatterers are needed, since then we have 9 available estimated parameters (1 AOA, 1 AOD, 1 TOA per path) and 9 unknowns (6 scalars for the scatterer locations $\mathbf{s}_k$, 3 scalars for $\mathbf{p}$ and $\alpha$).

\item \textbf{Unknown:} For the case that the receiver does not know whether it operates in \ac{NLOS} or \ac{OLOS}, %the \ac{LOS} probability can be used for different environments (e.g., open square, shopping mall, and office environment) as presented in \cite{Jarvelainen}. In this case, 
the receiver could apply the technique above under \ac{NLOS} and under \ac{OLOS}, separately. This will give two solutions with different cost (measured in terms of \eqref{EXIP4} and \eqref{EXIP4b}). The best solution (the one with lowest cost) can then be retained. 
\end{itemize}
%\subsection*{\textcolor{red}{Remark: Computational Analysis}}
The complexity analysis for each step of the aforementioned algorithm is presented in Appendix \ref{comprep}.
\section{Simulation Results}
In this section, we present simulation results show the values of the bounds and the performance of the proposed estimators for different parameters.

%to demonstrate the performance of the bound and the estimator with respect to different parameters. 
\subsection{Simulation Setup}
 We consider a scenario representative of indoor localization in a small conference room with the maximum distance between MS and BS of 4 meters \cite{Maltsevx}. We set $f_c = 60 \:\mathrm{GHz}$, $B=100 \:\mathrm{MHz}$, $c=0.299792\:\mathrm{m}/\mathrm{ns}$, and $N=20$. The geometry-based statistical path loss is used with path length $d_{k}$ and the number of reflectors in each path is set to one, i.e., it is assumed that there is one reflector in each \ac{NLOS} path \cite{geomted}. The path loss $\rho_{k}$ between \ac{BS} and \ac{MS} for the $k$-th path is computed based on geometry statistics \cite{Qian1,Qian2}. We set 
\begin{equation}\label{pathgeom1}
 1/\rho_{k}=\sigma^{2}_{0}\mathbb{P}_{0}(d_{k,2})\xi^{2}(d_{k})\left(\frac{\lambda_{c}}{4\pi d_{k}}\right)^{2},
 \end{equation}
where $\sigma^{2}_{0}$ is the reflection loss, $\mathbb{P}_{0}(d_{k,2})=(\gamma_{r}d_{k,2})^{2}e^{-\gamma_{r}d_{k,2}}$ denotes the Poisson distribution of environment geometry with density $\gamma_{r}$ (set to $1/7$ \cite{geomted}), $\xi^{2}(d_{k})$ denotes the atmospheric attenuation over distance $d_{k}$, and the last term is the free space path loss over distance $d_{k}$. For the \ac{LOS} link, we obtain
\begin{equation}\label{pathgeom2}
 1/\rho_{0}=\xi^{2}(d_{0})\left(\frac{\lambda_{c}}{4\pi d_{0}}\right)^{2}.
 \end{equation}
The average reflection loss for the first-order reflection $\sigma^{2}_{0}$ is set to $-10$ dB with the \ac{RMS} deviation equal to $4$ dB \cite{newref5gsix}, and the  atmospheric attenuation over distance $d_{k}$ is set to $16$ dB/km \cite{Rappaport}. The number of transmit and receive antennas are set to $N_{t}=65$ and $N_{r}=65$, respectively. The number of simultaneous beams is $M_{t}=1$, and the number of sequentially transmitted signals is $G=32$, unless otherwise stated. The \ac{BS} is located at $\mathbf{q}\:[\mathrm{m}]=[0, 0]^{\mathrm{T}}$ and the \ac{MS} is located at $\mathbf{p}\:[\mathrm{m}]=[4, 0]^{\mathrm{T}}$ with the rotation angle $\alpha=0.1\:\mathrm{rad}$. The elements of the analog beamformers are generated as random values uniformly distributed on the unit circle. The sequences $\tilde{\mathbf{x}}^{(g)}[n]=\mathbf{F}^{(g)}_{\mathrm{BB}}[n]\mathbf{x}^{(g)}[n]$ are obtained as complex exponential terms $e^{j\phi_{g,n}}$ with uniform random phases in $[0, 2\pi)$ along different subcarriers, indexed by $n$, and sequentially transmitted symbols, indexed by $g$.
%We set $\mathbf{x}^{(g)}[n]$ to be a random sequence ($M_t=1$) of G=32 sequentially transmitted QPSK symbols, while $\mathbf{F}^{(g)}[n]$  is set to have entries uniformly distributed on the unit circle. 
The values of the \ac{CRB} for $\sqrt{\mathrm{CRB}(\tau_{k})}$, $\sqrt{\mathrm{CRB}(\theta_{\mathrm{Rx},k})}$, and $\sqrt{\mathrm{CRB}(\theta_{\mathrm{Tx},k})}$ are defined similar to PEB and REB in \eqref{PEBexpr} and \eqref{REBexpr}, that is, by inverting the \ac{FIM} $\mathbf{J}_{\tilde{\boldsymbol{\eta}}}$ from \eqref{TFIM1}, choosing the corresponding diagonal entries and taking the square root. Finally, the received \acf{SNR} is defined as 
\begin{align}
\mathrm{SNR}\triangleq  \frac{\mathbb{E}[\Vert \mathrm{diag}\{\mathbf{\Omega}[0],\ldots \mathbf{\Omega}[N-1]\} \mathrm{vec}\{\check{\mathbf{h}}[0],\ldots,\check{\mathbf{h}}[N-1]\}\Vert^{2}_{2}]}{\mathbb{E}[\Vert   \mathrm{vec}\{\check{\mathbf{n}}[0],\ldots,\check{\mathbf{n}}[N-1]\} \Vert^{2}_{2}]},
\end{align}
in which $\mathrm{diag}\{\cdot \}$ creates a block diagonal matrix from its arguments and $\mathrm{vec}\{ \cdot\}$ creates a tall column vector from its arguments. 
%+\check{\mathbf{n}}[n],
%$\mathrm{SNR}\triangleq\mathbb{E}[\Vert\bar{\widetilde{\mathbf{\Omega}}}\bar{\mathbf{h}}\Vert^{2}_{2}]/\mathbb{E}[\Vert\bar{\mathbf{n}}\Vert^{2}_{2}]$ where $\bar{\mathbf{h}}=[\bar{\mathbf{h}}^{\mathrm{T}}[0],\ldots,\bar{\mathbf{h}}^{\mathrm{T}}[N-1]]^{\mathrm{T}}$ with $\bar{\mathbf{h}}[n]$ denoting the column-wise vector form of $\mathbf{H}[n]$, and $\bar{\widetilde{\mathbf{\Omega}}}=\mathrm{diag}\{\widetilde{\mathbf{\Omega}}[0],\ldots,\widetilde{\mathbf{\Omega}}[N-1]\}$ with $$\widetilde{\mathbf{\Omega}}[n]=\begin{bmatrix}\widetilde{\mathbf{\Omega}}^{(1)}[n]\\\vdots\\\widetilde{\mathbf{\Omega}}^{(G)}[n] \end{bmatrix},$$ and $\widetilde{\mathbf{\Omega}}^{(g)}[n]=(\mathbf{F}^{(g)}[n]\mathbf{x}^{(g)}[n])^{\mathrm{T}}\otimes\mathbf{I}_{N_{r}}$.

The performance of the \ac{RMSE} of the estimation algorithm was assessed from $1000$ Monte Carlo realizations. The false alarm probability was set to ${P}_{\mathrm{fa}}=10^{-3}$ to determine the threshold $\delta$. 

 \subsection{Results and Discussion}  \label{results}
%\subsubsection*{\textcolor{blue}{The Performance of the PEB vs $G$}}
\subsubsection*{The Performance versus number of sequential beams}
\begin{figure}   
\centering
\psfrag{G}[c][]{\footnotesize number of beams}
\psfrag{PEB}[c][]{\scriptsize PEB}
\psfrag{RMSE}[c][]{\scriptsize \qquad$\mathrm{RMSE}(\hat{\mathbf{p}})$}
\psfrag{PEB,G=24}[c][]{\scriptsize \qquad PEB, $24$ beams}
\psfrag{RMSE, G=4}[c][]{\scriptsize \:\:\:\qquad$\mathrm{RMSE}(\hat{\mathbf{p}})$, $4$ beams}
\psfrag{RMSE, G=8}[c][]{\scriptsize \:\:\:\qquad$\mathrm{RMSE}(\hat{\mathbf{p}})$, $8$ beams}
\psfrag{RMSE, G=20}[c][]{\scriptsize \qquad$\mathrm{RMSE}(\hat{\mathbf{p}})$, $20$ beams}
\psfrag{RMSE, G=24}[c][]{\scriptsize \qquad$\mathrm{RMSE}(\hat{\mathbf{p}})$, $24$ beams}
\psfrag{RMSE, G=26}[c][]{\scriptsize \qquad$\mathrm{RMSE}(\hat{\mathbf{p}})$, $26$ beams}
\psfrag{CDF}[c][]{\footnotesize CDF}
\psfrag{Localization error [m]}[c][]{\footnotesize Localization error [m]}
\includegraphics[width=0.7\columnwidth]{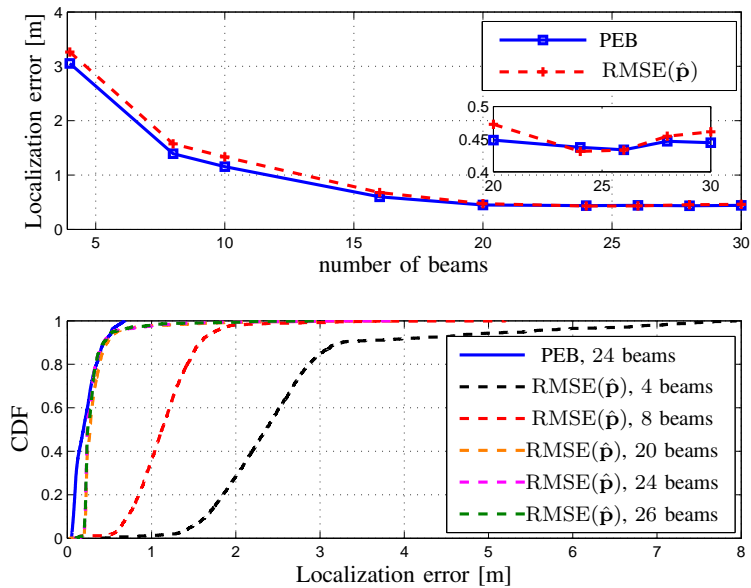}
\caption{The effect of increasing the number of beams on (top) PEB and $\mathrm{RMSE}(\hat{\mathbf{p}})$ at $\mathrm{CDF}=0.9$ and (bottom) CDF plots for LOS conditions.}
  \label{Geffect}
\end{figure}

%The beams are generated using the random phase shifters with the entries uniformly distributed on the unit circle.The random sequence $\tilde{\mathbf{x}}^{(g)}[n]=\mathbf{F}^{(g)}_{\mathrm{BB}}[n]\mathbf{x}^{(g)}[n]$ follows complex exponential terms $e^{j\phi_{g,n}}$ with random phases along different subcarriers $n$ and symbols $g$ for $M_{t}=1$. 
Fig. \ref{Geffect} shows the \ac{CDF} of the PEB and the $\mathrm{RMSE}(\hat{\mathbf{p}})$ as a function of the number of beams for LOS conditions. The MS can be anywhere in a rectangle with vertices at the coordinates (in meters): $(2,0)$, $(4,0)$, $(2,0.3)$, and $(4,0.3)$. %The values at the  $90\%$ percentile are also plotted. based on \ac{CDF} at $90\%$ with respect to the number of beams for LOS and the same total transmit power.
The signal is scaled so that the total transmit power is kept constant.
%To generate the beams, first we define the quantized angle parameters as
%\begin{equation}\label{quanangl1}
%\mathcal{G}=\left\{\phi_{g}: \phi_{g}\in [-\pi/2, \pi/2], g=1,\ldots,G\right\}.
%\end{equation}
%The angle parameters $\{\phi_{g}\}$ in \eqref{quanangl1} are defined so that the terms $\{\sin(\phi_{g})\}$ in the definition of array response $\mathbf{a}(\phi_{g})$ are uniformly distributed in $[-1, 1]$.
%It is clear that after transmitting 20 beams, the PEB saturates and increasing the number of beams does not improve the localization accuracy noticeably. 
%\textcolor{blue}{After transmitting sufficient number of randomly selected beams the localization accuracy is approximately the same with $\mathrm{CDF}=0.9$.}
By increasing the number of beams $G$, the probability of covering the target location in the specified area with a certain accuracy increases. In other words, due to the ergodicity of the process localization accuracy with a certain number of randomly selected sequential beams in each step converges to a constant value for sufficient number of beams $G$. The reason is that for a larger number of beams, the bound decreases thanks to the better spatial coverage. But this effect vanished when the number of beams is sufficient to cover the area where the \ac{MS} may be located, and then increasing the number of beams only translated into an increased complexity. In principle, the 3 dB beam width for the ULA is approximately $2/N_{t}$, thus reducing when increasing the number of transmit antennas $N_{t}$. Consequently, the number of required beams $G$ to cover the target location in the specified area with the same probability increases. Similarly, by reducing the number of transmit antennas $N_{t}$, the number of required beams $G$ to cover the area decreases. However, the localization accuracy is improved for the case with larger number of transmit antennas $N_{t}$ with the cost of transmitting more beams $G$ for the same coverage. It is observed that for the aforementioned system parameters, $G\geq 20$ randomly selected beams approximately provides the same localization accuracy with $\mathrm{CDF}=0.9$. Note that fewer beams would be needed under a well-chosen deterministic strategy.
%\textcolor{blue}{Moreover, the 3 dB beam width for the ULA in the BS is approximately $2/N_{t}$ that is reduced by increasing the number of antenna elements $N_{t}$. This leads to requiring more number of randomly selected beams to cover the target location with approximately the same accuracy. Similarly, using smaller number of antenna elements $N_{t}$ the coverage probability with smaller number of beams $G$ is increased with the cost of reduced localization accuracy. It is observed that for the simulation parameters $G\geq 20$ randomly selected beams approximately provides the same localization accuracy with $\mathrm{CDF}=0.9$.} %Moreover, the $\mathrm{RMSE}(\hat{\mathbf{p}})$ does not change after $G=20$ and approaches the PEB in the saturation region at $\mathrm{CDF}=0.9$. Consequently, increasing the number of beams only increases the complexity and the PEB and $\mathrm{RMSE}(\hat{\mathbf{p}})$ do not change considerably. 
%This is independent from being in LOS or NLOS conditions. So, we set $G=32$ for coverage in the entire simulations.
The same behavior has been observed in NLOS conditions. 
 \subsubsection*{Performance in LOS}
\begin{figure}   
\centering
\psfrag{snr2=-20dB}[c][]{\footnotesize \qquad\qquad$\mathrm{SNR}=-20$ dB}
\psfrag{snr3=-10dB}[c][]{\footnotesize \qquad\qquad$\mathrm{SNR}=-10$ dB}
\psfrag{snr3=0dB}[c][]{\footnotesize \qquad\qquad$\mathrm{SNR}=0$ dB}
%\psfrag{true}[c][]{\footnotesize true}
%\psfrag{AOA1}[c][]{\small $\pi-\hat{\theta}_{\mathrm{Rx},0}\:[\mathrm{rad}]$}
%\psfrag{AOD1}[c][]{\small $\hat{\theta}_{\mathrm{Tx},0}\:[\mathrm{rad}]$}
%\psfrag{TOA1}[c][]{\small $\hat{\tau}_{0}\:[\mathrm{ns}]$}
%\psfrag{TOA0 at −20dB}[c][]{\footnotesize $\mathrm{TOA}_{0}$ at −20 dB}
%\psfrag{TOA0 at −10dB}[c][]{\footnotesize $\mathrm{TOA}_{0}$ at −10 dB}
%\psfrag{TOA0 at 0dB}[c][]{\footnotesize $\mathrm{TOA}_{0}$ at 0 dB}
%\psfrag{AOA0 at −20dB}[c][]{\footnotesize $\mathrm{AOA}_{0}$ at −20 dB}
%\psfrag{AOA0 at −10dB}[c][]{\footnotesize $\mathrm{AOA}_{0}$ at −10 dB}
%\psfrag{AOA0 at 0dB}[c][]{\footnotesize $\mathrm{AOA}_{0}$ at 0 dB}
%\psfrag{AOD0 at −20dB}[c][]{\footnotesize $\mathrm{TOA}_{0}$ at −20 dB}
%\psfrag{AOD0 at −10dB}[c][]{\footnotesize $\mathrm{TOA}_{0}$ at −10 dB}
%\psfrag{AOD0 at 0dB}[c][]{\footnotesize $\mathrm{TOA}_{0}$ at 0 dB}
\psfrag{TOA0 at x}[c][]{\footnotesize \qquad\qquad$\:\:\mathrm{TOA}_{0}$ at $-20$ dB}
\psfrag{TOA0 at y}[c][]{\footnotesize \qquad\qquad$\:\:\mathrm{TOA}_{0}$ at $-10$ dB}
\psfrag{TOA0 at z}[c][]{\footnotesize \qquad\qquad$\:\:\mathrm{TOA}_{0}$ at $0$ dB}
\psfrag{AOA0 at x}[c][]{\footnotesize \qquad\qquad$\:\:\mathrm{AOA}_{0}$ at $-20$ dB}
\psfrag{AOA0 at y}[c][]{\footnotesize \qquad\qquad$\:\:\mathrm{AOA}_{0}$ at $-10$ dB}
\psfrag{AOA0 at z}[c][]{\footnotesize \qquad\qquad$\:\:\mathrm{AOA}_{0}$ at $0$ dB}
\psfrag{AOD0 at x}[c][]{\footnotesize \qquad\qquad$\:\:\mathrm{AOD}_{0}$ at $-20$ dB}
\psfrag{AOD0 at y}[c][]{\footnotesize \qquad\qquad$\:\:\mathrm{AOD}_{0}$ at $-10$ dB}
\psfrag{AOD0 at z}[c][]{\footnotesize \qquad\qquad$\:\:\mathrm{AOD}_{0}$ at $0$ dB}
\psfrag{RMSETOA1}[c][]{\footnotesize $\mathrm{RMSE}(\hat{\tau}_{0})$ [ns]}
\psfrag{RMSEAOA1}[c][]{\footnotesize $\mathrm{RMSE}(\hat{\theta}_{\mathrm{Rx},0})$ [rad]}
\psfrag{RMSEAOD1}[c][]{\footnotesize $\mathrm{RMSE}(\hat{\theta}_{\mathrm{Tx},0})$ [rad]}
\psfrag{iteration index}[c][]{\footnotesize iteration index}
\includegraphics[width=0.58\columnwidth]{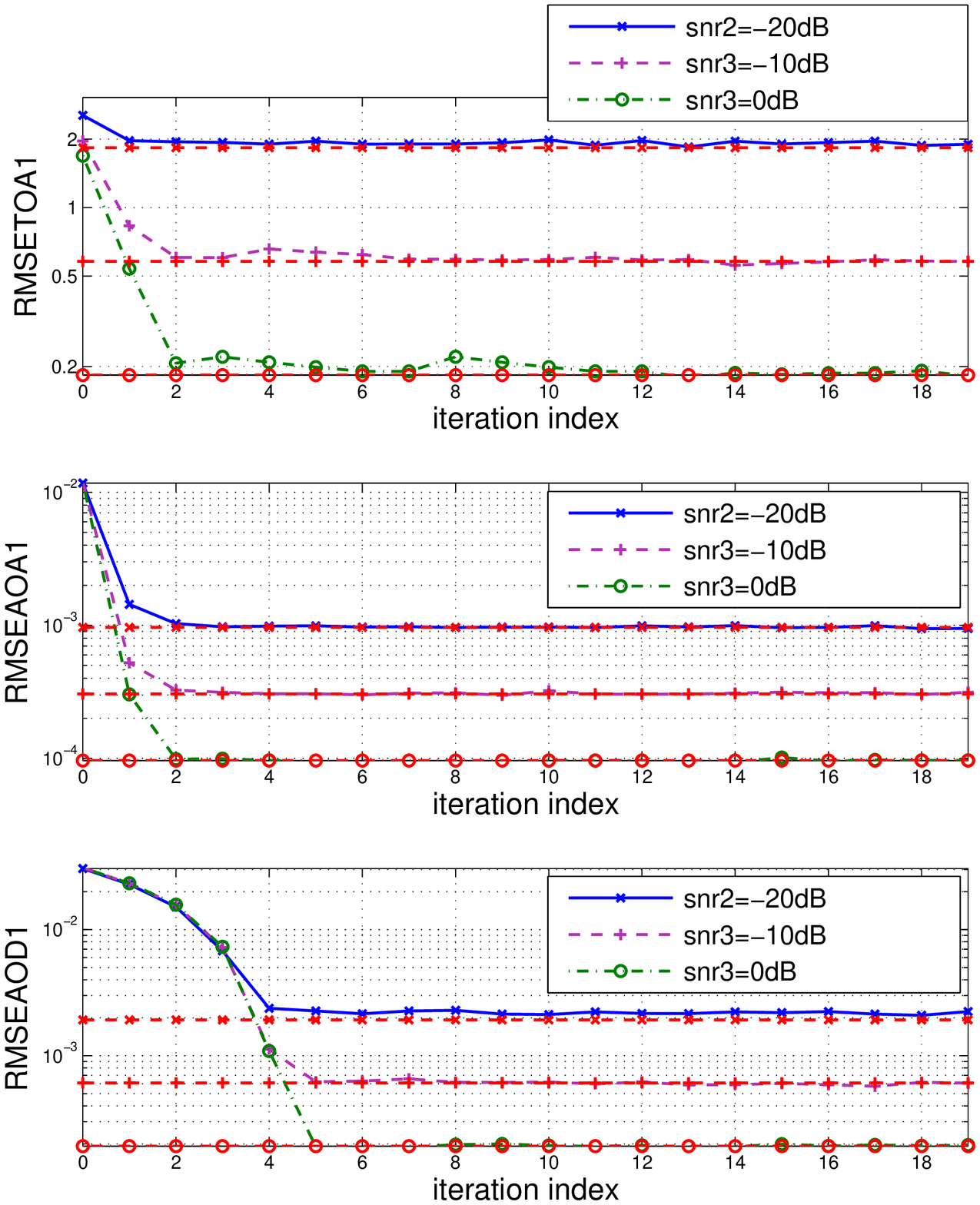}
\caption{The evolution of \ac{RMSE} of TOA and AOA/AOD for the LOS for $\mathrm{SNR}=-20\:\mathrm{dB}, -10\:\mathrm{dB}, 0\:\mathrm{dB}$. The red lines with the same markers show the bounds for the same value of \ac{SNR} corresponding to the \ac{RMSE} of TOA and AOA/AOD.}
  \label{ParamvsIter}
\end{figure}

\begin{figure}   
\centering
\psfrag{RMSE0}[c][]{\footnotesize $\mathrm{RMSE}(\hat{\tau}_{0})$ [ns]}
\psfrag{RMSE1}[c][]{\footnotesize $\mathrm{RMSE}(\hat{\theta}_{\mathrm{Rx},0})$ [rad]}
\psfrag{RMSE2}[c][]{\footnotesize $\mathrm{RMSE}(\hat{\theta}_{\mathrm{Tx},0})$ [rad]}
\psfrag{SNR}[c][]{\footnotesize SNR (in dB)}
\psfrag{RMSEx1}[c][]{\tiny \qquad$\:\:\mathrm{RMSE}(\hat{\tau}_{0})$}
\psfrag{RMSEx2}[c][]{\tiny \qquad$\:\:\mathrm{RMSE}(\hat{\theta}_{\mathrm{Rx},0})$}
\psfrag{RMSEx3}[c][]{\tiny \qquad$\:\:\mathrm{RMSE}(\hat{\theta}_{\mathrm{Tx},0})$}
\psfrag{RMSEy1}[c][]{\tiny \qquad$\:\:\sqrt{\mathrm{CRB}(\tau_{0})}$}
\psfrag{RMSEy2}[c][]{\tiny \qquad$\:\:\sqrt{\mathrm{CRB}(\theta_{\mathrm{Rx},0})}$}
\psfrag{RMSEy3}[c][]{\tiny \qquad$\:\:\sqrt{\mathrm{CRB}(\theta_{\mathrm{Tx},0})}$}
\psfrag{iteration index}[c][]{\small iteration index}
\includegraphics[width=0.58\columnwidth]{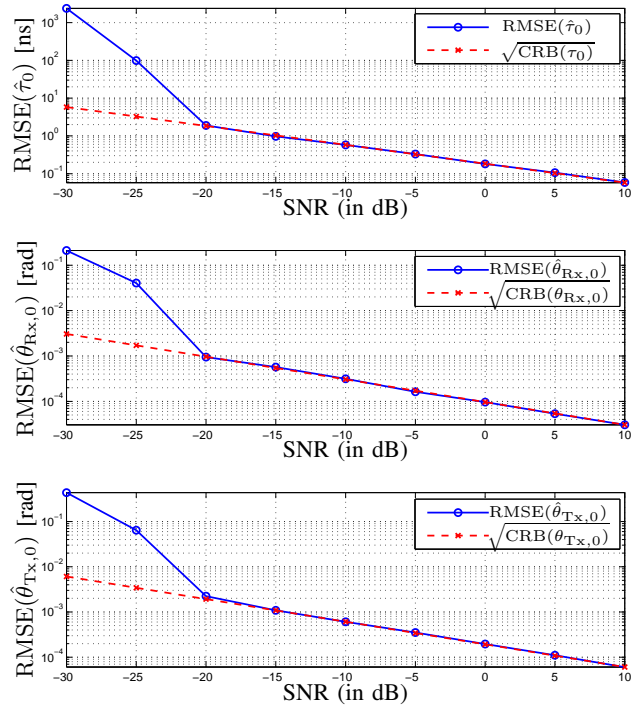}
\caption{RMSE in dB scale plotted against received SNR for TOA and AOA/AOD in the LOS conditions. The red lines show the corresponding bounds.}
  \label{ParamvsSNR}
\end{figure}

\begin{figure}   
\centering
\psfrag{SNR}[c][]{\footnotesize SNR (in dB)}
\psfrag{REBb}[c][]{\footnotesize \qquad REB}
\psfrag{PEBb}[c][]{\footnotesize \qquad PEB}
\psfrag{RMSExx}[c][]{\footnotesize $\mathrm{RMSE}(\hat{\alpha})$ [rad]}
\psfrag{RMSEyy}[c][]{\footnotesize $\mathrm{RMSE}(\hat{\mathbf{p}})$ [m]}
\psfrag{RMSEx}[c][]{\footnotesize \qquad$\mathrm{RMSE}(\hat{\alpha})$}
\psfrag{RMSEy}[c][]{\footnotesize \qquad$\mathrm{RMSE}(\hat{\mathbf{p}})$}
\includegraphics[width=0.57\columnwidth]{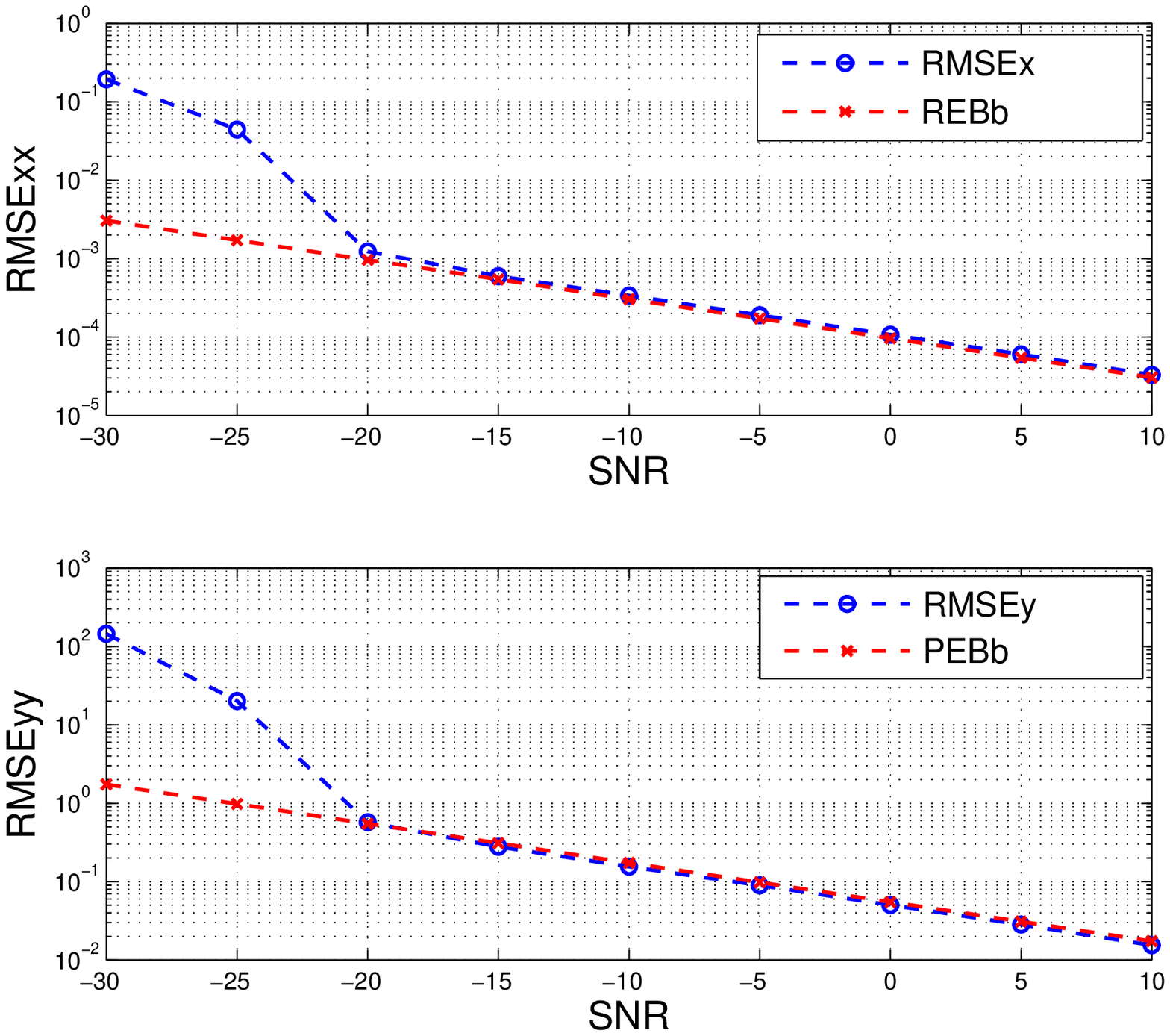}
\caption{RMSE in dB scale plotted against received SNR for rotation angle (top) and position (bottom) in the \ac{LOS}. The red lines show the corresponding bounds.}
  \label{PREBvsSNR_losss}
\end{figure}

Fig.~\ref{ParamvsIter} shows the evolution of the \ac{RMSE} of TOA and AOA/AOD in the LOS conditions. The Cram\'{e}r-Rao bounds are shown by the red lines with the corresponding markers. It is observed that after a few iterations of Algorithm 2 the \ac{RMSE} of TOA and AOA/AOD converges to the corresponding bounds even for $\mathrm{SNR}=-20\:\mathrm{dB}, -10\:\mathrm{dB}, 0\:\mathrm{dB}$.  The performance of the \ac{RMSE} of the estimation algorithm  with respect to different values of the received SNR is shown in Fig.~\ref{ParamvsSNR}--\ref{PREBvsSNR_losss}. 
It is observed that after $\mathrm{SNR}\approx -20$ dB the \ac{RMSE} of the TOA, AOA/AOD, rotation angle, and position converge to their corresponding bounds (red dashed lines). Moreover, the proposed algorithm performs well even for very low values of the received SNR, which is the typical case at \ac{mm-wave} systems before beamforming. 
We observe that at $\mathrm{SNR}\approx -20\:\mathrm{dB}$ the TOA, AOA/AOD, rotation angle, and position approach the corresponding bounds.


\subsubsection*{Performance in NLOS}
\begin{figure}   
\centering
%\psfrag{snr1=-20dB}[c][]{\footnotesize $\mathrm{SNR}=-20$ dB}
\psfrag{snr2=-5dB}[c][]{\footnotesize $\qquad\mathrm{SNR}=-5$ dB}
\psfrag{snr3=0dB}[c][]{\footnotesize $\qquad\mathrm{SNR}=0$ dB}
\psfrag{RMSETOA1}[c][]{\footnotesize $\mathrm{RMSE}(\hat{\tau}_{0})$ [ns]}
\psfrag{RMSETOA2}[c][]{\footnotesize $\mathrm{RMSE}(\hat{\tau}_{1})$ [ns]}
\psfrag{RMSEAOA1}[c][]{\footnotesize $\mathrm{RMSE}(\hat{\theta}_{\mathrm{Rx},0})$ [rad]}
\psfrag{RMSEAOA2}[c][]{\footnotesize $\mathrm{RMSE}(\hat{\theta}_{\mathrm{Rx},1})$ [rad]}
\psfrag{RMSEAOD1}[c][]{\footnotesize $\mathrm{RMSE}(\hat{\theta}_{\mathrm{Tx},0})$ [rad]}
\psfrag{RMSEAOD2}[c][]{\footnotesize $\mathrm{RMSE}(\hat{\theta}_{\mathrm{Tx},1})$ [rad]}
\psfrag{iteration index}[c][]{\footnotesize iteration index}
\includegraphics[width=0.9\columnwidth]{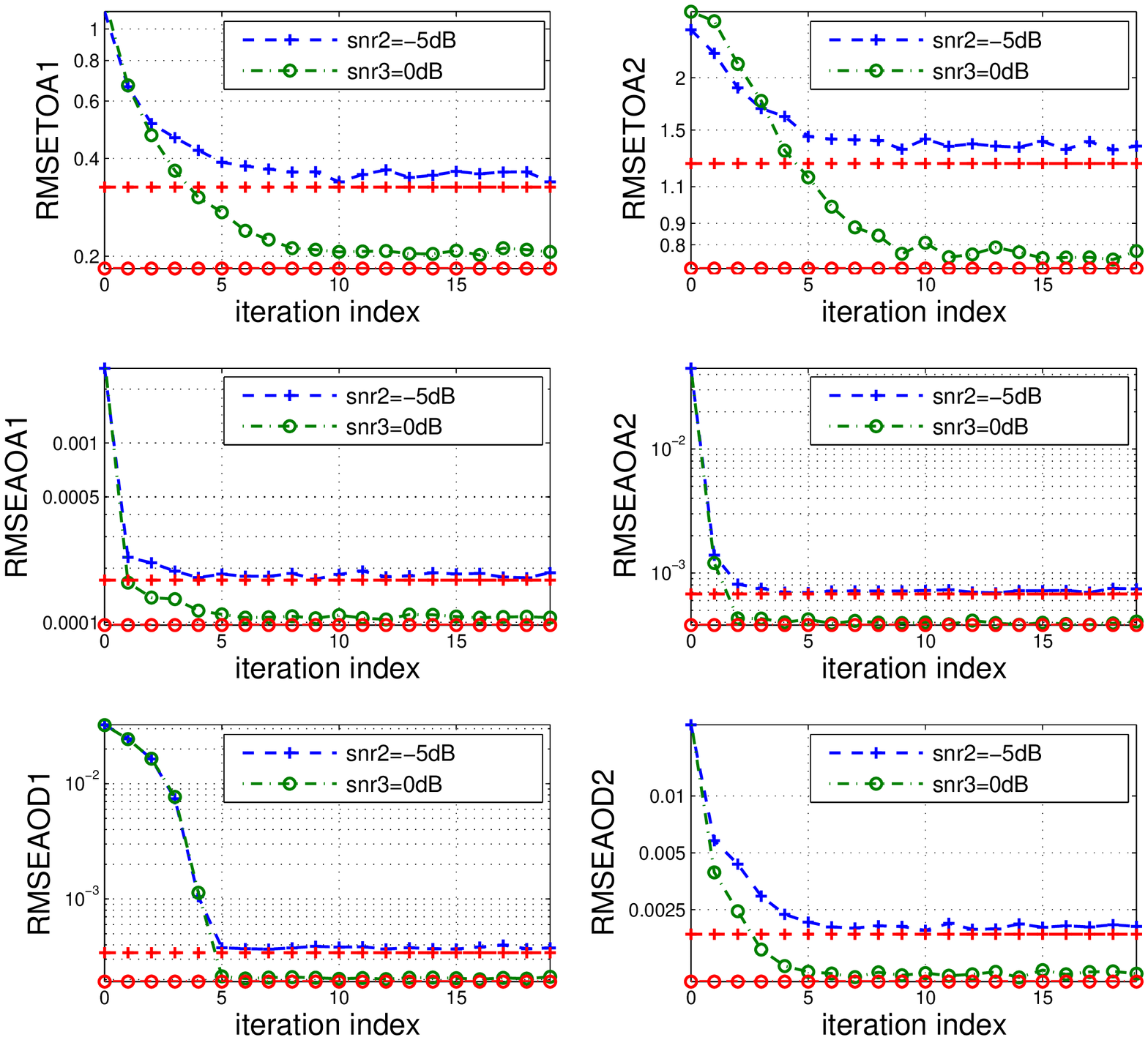}
\caption{The evolution of \ac{RMSE} of TOA and AOA/AOD for the LOS (left column) and the NLOS (right column) paths at $\mathrm{SNR}=-5\:\mathrm{dB}, 0\:\mathrm{dB}$. The red lines with the same markers show the bounds.}
  \label{ParamvsIterNLOS}
\end{figure}

Fig.~\ref{ParamvsIterNLOS} shows the evolution of the \ac{RMSE} of TOA and AOA/AOD for $1000$ Monte Carlo realizations in the presence of a scatterer located at $\mathbf{s}_{k}\:[\mathrm{m}]=[1.5, 0.4]^{\mathrm{T}}$. It can be observed that the \ac{RMSE} of the TOA and the AOA/AOD obtained with the proposed algorithm for both the parameters of the LOS and the reflected signals  converges to the theoretical also in this case, even at very low received SNR.    %  The performance of the \ac{RMSE} of the estimation algorithm with respect to the received \ac{SNR} is shown in Fig.~\ref{ParamvsSNRnlos}-\ref{PREBvsSNR}. It is observed that after a few iterations the \ac{RMSE} of TOA, AOA/AOD, rotation angle, and position converge to the corresponding bounds (shown by the red lines) even for very low values of the received SNR. 
At $\mathrm{SNR}\approx -5\:\mathrm{dB}$ the TOA, AOA/AOD, rotation angle, and position approach the corresponding bounds.
\begin{figure}   
\centering
\psfrag{RMSE0}[c][]{\footnotesize $\mathrm{RMSE}(\hat{\tau}_{k})$ [ns]}
\psfrag{RMSE1}[c][]{\footnotesize $\mathrm{RMSE}(\hat{\theta}_{\mathrm{Rx},k})$ [rad]}
\psfrag{RMSE2}[c][]{\footnotesize $\mathrm{RMSE}(\hat{\theta}_{\mathrm{Tx},k})$ [rad]}
\psfrag{SNR}[c][]{\footnotesize SNR (in dB)}
\psfrag{RMSEx1}[c][]{\footnotesize \:$\qquad\mathrm{RMSE}(\hat{\tau}_{0})$}
\psfrag{RMSEy1}[c][]{\footnotesize \:$\qquad\mathrm{RMSE}(\hat{\tau}_{1})$}
\psfrag{RMSEx2}[c][]{\footnotesize \:\:$\qquad\mathrm{RMSE}(\hat{\theta}_{\mathrm{Rx},0})$}
\psfrag{RMSEy2}[c][]{\footnotesize \:\:$\qquad\mathrm{RMSE}(\hat{\theta}_{\mathrm{Rx},1})$}
\psfrag{RMSEx3}[c][]{\footnotesize \:\:$\qquad\mathrm{RMSE}(\hat{\theta}_{\mathrm{Tx},0})$}
\psfrag{RMSEy3}[c][]{\footnotesize \:\:$\qquad\mathrm{RMSE}(\hat{\theta}_{\mathrm{Tx},1})$}
\psfrag{TOA0}[c][]{\footnotesize \:$\qquad\sqrt{\mathrm{CRB}(\tau_{0})}$}
\psfrag{TOA1}[c][]{\footnotesize \:$\qquad\sqrt{\mathrm{CRB}(\tau_{1})}$}
\psfrag{AOA0}[c][]{\footnotesize \:\:$\qquad\sqrt{\mathrm{CRB}(\theta_{\mathrm{Rx},0})}$}
\psfrag{AOA1}[c][]{\footnotesize \:\:$\qquad\sqrt{\mathrm{CRB}(\theta_{\mathrm{Rx},1})}$}
\psfrag{AOD0}[c][]{\footnotesize \:\:$\qquad\sqrt{\mathrm{CRB}(\theta_{\mathrm{Tx},0})}$}
\psfrag{AOD1}[c][]{\footnotesize \:\:$\qquad\sqrt{\mathrm{CRB}(\theta_{\mathrm{Tx},1})}$}
\includegraphics[width=0.7\columnwidth]{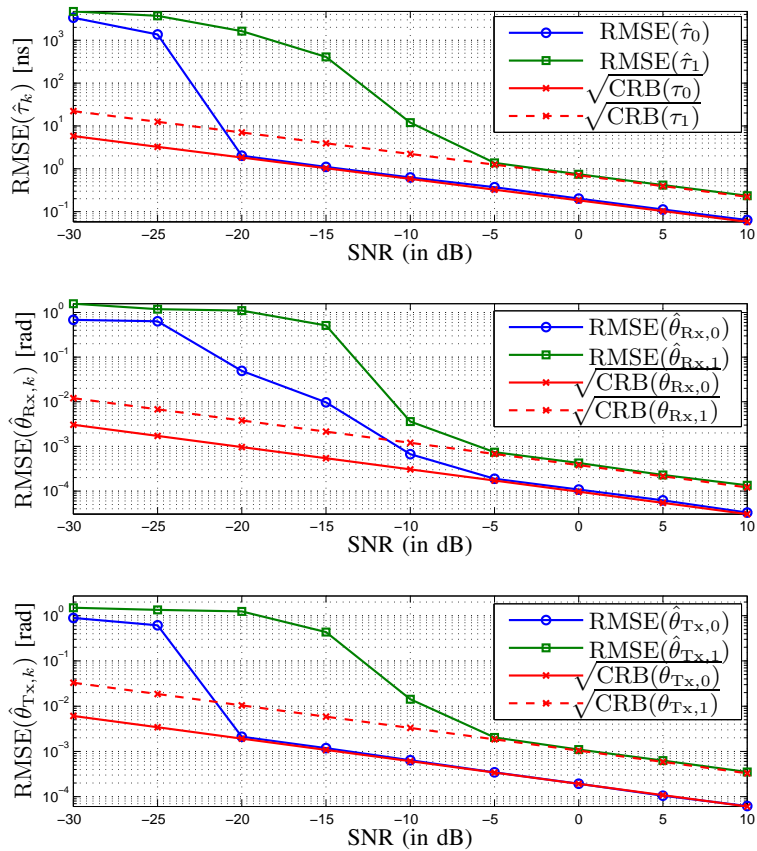}
\caption{RMSE in dB scale for the \ac{NLOS} plotted against received SNR for TOA and AOA/AOD in the presence of a scatterer located at $\mathbf{s}_{k}\:[\mathrm{m}]=[1.5, 0.4]^{\mathrm{T}}$. The red lines show the corresponding bounds.}
  \label{ParamvsSNRnlos}
\end{figure}
\begin{figure}   
\centering
\psfrag{SNR}[c][]{\footnotesize SNR (in dB)}
\psfrag{REBb}[c][]{\footnotesize REB}
\psfrag{PEBb}[c][]{\footnotesize PEB}
\psfrag{RMSExx}[c][]{\footnotesize $\mathrm{RMSE}(\hat{\alpha})$ [rad]}
\psfrag{RMSEyy}[c][]{\footnotesize $\mathrm{RMSE}(\hat{\mathbf{p}})$ [m]}
\psfrag{RMSEx}[c][]{\footnotesize $\qquad\mathrm{RMSE}(\hat{\alpha})$}
\psfrag{RMSEy}[c][]{\footnotesize $\qquad\mathrm{RMSE}(\hat{\mathbf{p}})$}
\includegraphics[width=0.57\columnwidth]{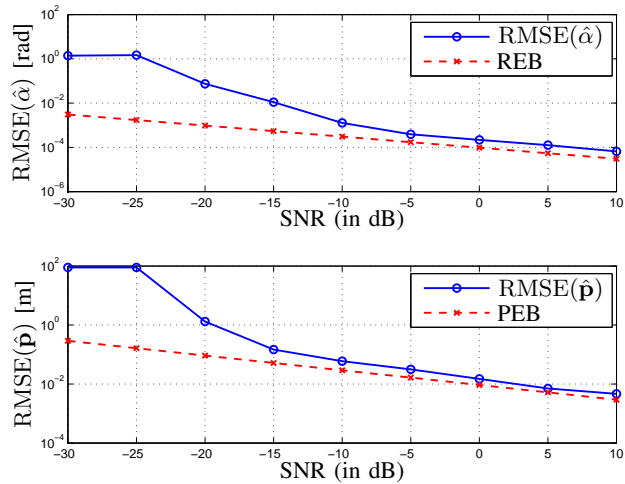}
\caption{RMSE in dB scale for the \ac{NLOS} plotted against received SNR for rotation angle (top) and position (bottom) in the presence of a scatterer located at $\mathbf{s}_{k}\:[\mathrm{m}]=[1.5, 0.4]^{\mathrm{T}}$. The red lines show the corresponding bounds.}
  \label{PREBvsSNR}
\end{figure}
\begin{figure}   
\centering
\psfrag{SNR}[c][]{\small SNR (in dB)}
\psfrag{REBb}[c][]{\scriptsize REB}
\psfrag{PEBb}[c][]{\scriptsize PEB}
\psfrag{RMSEa}[c][]{\small $\mathrm{RMSE}(\hat{\alpha})$ [rad]}
\psfrag{RMSEp}[c][]{\small $\mathrm{RMSE}(\hat{\mathbf{p}})$ [m]}
\psfrag{RMSEy}[c][]{\scriptsize $\qquad\qquad\qquad\qquad\mathrm{RMSE}(\hat{\alpha}),\Delta\alpha\:[\mathrm{rad}]=0.01$}
\psfrag{RMSEyy}[c][]{\scriptsize $\qquad\qquad\qquad\qquad\mathrm{RMSE}(\hat{\alpha}),\Delta\alpha\:[\mathrm{rad}]=0.05$}
\psfrag{RMSEx}[c][]{\scriptsize $\qquad\qquad\qquad\qquad\mathrm{RMSE}(\hat{\mathbf{p}}),\Delta\alpha\:[\mathrm{rad}]=0.01$}
\psfrag{RMSExx}[c][]{\scriptsize $\qquad\qquad\qquad\qquad\mathrm{RMSE}(\hat{\mathbf{p}}),\Delta\alpha\:[\mathrm{rad}]=0.05$}
\includegraphics[width=0.57\columnwidth]{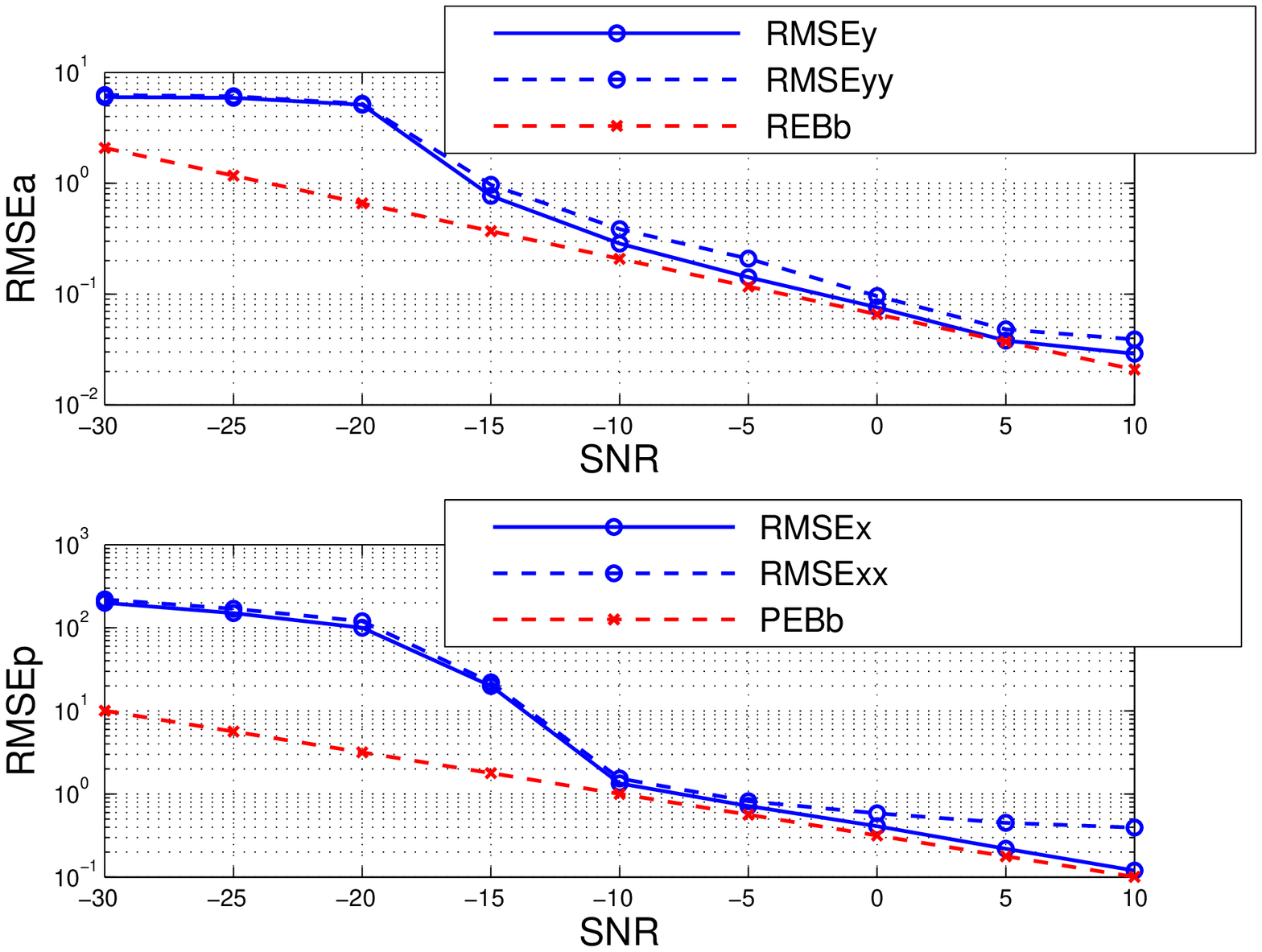}
\caption{RMSE in dB scale plotted against received SNR for rotation angle (top) and position (bottom) in the \ac{OLOS} with three scatterers located at $\mathbf{s}_{k}\:[\mathrm{m}]=[1.5, 0.4+0.5(k-1)]^{\mathrm{T}}$ for $k=1, 2, 3$ and $\Delta\alpha\:[\mathrm{rad}]=\{0.01,0.05\}$. The red lines show the corresponding bounds.}
 \label{PREBvsSNRolos}
\end{figure}

\subsubsection*{Performance in OLOS}
Finally, the performance in the \ac{OLOS} case for three scatterers located at $\mathbf{s}_{k}\:[\mathrm{m}]=[1.5, 0.4+0.5(k-1)]^{\mathrm{T}}$ for $k=1, 2, 3$ is investigated in this section using two different initializations of the rotation angle: one with grid resolution $\Delta\alpha\:[\mathrm{rad}]=0.01$ and one with $\Delta\alpha\:[\mathrm{rad}]=0.05$. For both, we set 
 %that generate the lowest possible \ac{RMSE} and are chosen from the grid of points with the resolutions $\Delta\alpha\:[\mathrm{rad}]=\{0.01,0.05\}$ and
  $\alpha_{m}\:[\mathrm{rad}]=0.5$. Fig.~\ref{PREBvsSNRolos} shows the performance of the \ac{RMSE}  with respect to the received \ac{SNR} for position and rotation angle estimation. The proposed estimation method approaches the bound even for the initialization with the resolution $\Delta\alpha\:[\mathrm{rad}]=0.05$. However, the performance of the estimation algorithm is dependent on the resolution of the grid of points $\Delta\alpha$. In particular, a finer grid for the rotation angle leads to better initial estimates and thus a lower final RMSE. 
%the performance is improved for the rotation angle chosen from an interval with the finer grid of points that leads to the initialization of the parameters for solving the nonlinear least squares problem in \eqref{EXIP4b} close to the optimal values using the \ac{LMA} algorithm.
  For  $\mathrm{SNR}\approx -10\:\mathrm{dB}$ the \ac{RMSE} of position and rotation angle approach the corresponding bounds. We note that the OLOS values, for a fixed SNR, are significantly higher in the OLOS than in the NLOS case. 
\subsubsection*{Unknown Conditions}
To analyze the application of the algorithm when the propagation conditions are unknown, we consider the case where there are three scatterers and the LOS path is blocked, that is, the OLOS condition. Starting with the wrong assumption that the path with the shortest delay is the LOS path (i.e., the NLOS condition) leads to very large values of the cost function \eqref{EXIP4} compared to the actual value of the cost function \eqref{EXIP4b}. The results are summarized in Table. \ref{tab:assumptions} for the average value of the ratio ${\Delta v}\triangleq v_{\mathrm{nlos}}(\hat{\tilde{\boldsymbol{\eta}}})/v_{\mathrm{olos}}(\hat{\tilde{\boldsymbol{\eta}}}_{\mathrm{olos}})$ between the cost function with the wrong and true assumptions. The values in Table \ref{tab:assumptions} are obtained by averaging 100 realizations, and with a grid resolution of $\Delta\alpha=0.05 [\mathrm{rad}]$. The slight difference in the ratio for different values of SNR is due to the limited number of trials.
\begin{table}[!t]
\centering
%\begin{centering}
\caption{Unknown conditions} \label{tab:assumptions}
\begin{tabular}{|c|c|c|c|c|}
\hline 
 SNR (in dB) & -20 & -10 & 0 &10 
\\\hline 
${\Delta v}$& 5.5&5.2&5&5.3\\\hline 
\end{tabular}
%\end{centering}
\end{table}
It is clear that using the wrong assumption about the path with the shortest delay leads to much larger values of the cost function, i.e., the mean value of the ratio ${\Delta v}$ between the cost function with the wrong and true assumptions is on the order of $5$. The main reason for the increase of the cost function using the wrong assumption about the shortest path is that the estimate of \ac{MS} rotation angle obtained from the AOA and AOD of this path is heavily erroneous. When the shortest path is considered to be a LOS but it is really a reflection, there is a clear mismatch between the geometry of the propagation and the model equations, since there is a scatterer that breaks the direct relation between AOA and AOD existing with the LOS. This mismatch causes a large error in the initial position that is propagated to the final solution. Therefore, observing the ratio of cost functions, we can identify that the path with the shortest delay is related to the scatterer and the LOS path does not exist, that is to say, the OLOS condition is correctly recognized.

\subsubsection*{Comparison of LOS versus NLOS Performance}
\begin{figure}   
\centering
\psfrag{CDF}[c][]{\small CDF}
\psfrag{Localization error [m]}[c][]{\small Localization error [m]}
\psfrag{NLOS with 1 scatterer}[c][]{\footnotesize \qquad NLOS with 1 scatterer}
\psfrag{NLOS with 2 scatterers}[c][]{\footnotesize \qquad NLOS with 2 scatterers}
\psfrag{LOS}[c][]{\footnotesize \qquad LOS}
\includegraphics[width=0.57\columnwidth]{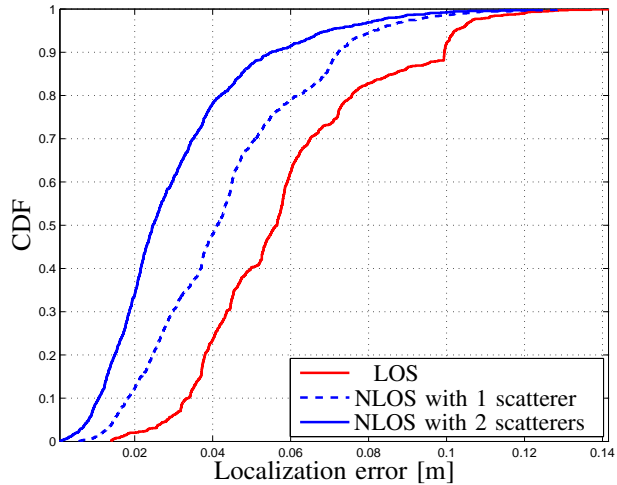}
\caption{CDF of the localization error in LOS and NLOS with one and two scatterers for $\mathrm{SNR}=-5$ dB and $G=20$.}
  \label{CDFdiff}
\end{figure}

Fig. \ref{CDFdiff} compares the performance of the positioning algorithm in LOS and NLOS for $\mathrm{SNR}= -5$ dB and $G=20$. The MS is anywhere in the same rectangle described at the beginning of Sec.~\ref{results}. The scatterers are located at coordinates (in meters) $\mathbf{s}_{1}=(1.5,0.4)$ and $\mathbf{s}_{2}=(1.5,0.6)$. The accuracy and robustness of the localization algorithm is improved by adding the scatterers compared to the case when only LOS is used. Moreover, the performance in the OLOS is much worse than in LOS or NLOS due to the severe effect of path loss as shown already in the paper by comparing Figs. \ref{PREBvsSNR_losss} and \ref{PREBvsSNR} with Fig. \ref{PREBvsSNRolos}.
%\begin{figure}   
%\centering
%%\includegraphics[width=1\columnwidth]{PEBHWvJune28.eps}
%\psfrag{p}[c][]{\small $\mathbf{p}$}
%\psfrag{q (0,0) m}[c][]{\small $\mathbf{q}\:(0,0)$ m}
%\psfrag{s1 (1.5,0.4) m}[c][]{\small $\mathbf{s}_{1}\:(1.5,0.4)$ m}
%\psfrag{s2 (1.5,0.6) m}[c][]{\small $\mathbf{s}_{2}\:(1.5,0.6)$ m}
%\psfrag{(2,0) m}[c][]{\small $(2,0)$ m}
%\psfrag{(4,0) m}[c][]{\small $(4,0)$ m}
%\psfrag{(4,0.3) m}[c][]{\small $(4,0.3)$ m}
%\psfrag{(2,0.3) m}[c][]{\small $(2,0.3)$ m}
%\includegraphics[width=0.8\columnwidth]{simsetup.eps}
%\caption{\textcolor{blue}{Simulation setup for the performance comparison in LOS and NLOS.}}
%  \label{simset}
%\end{figure}

\section{Conclusion}\label{SEC:Conclusion}

We have studied the determination of a receiver position and orientation using a single transmitter in a MIMO system. Our study includes \ac{LOS}, as well as \ac{NLOS} and \ac{OLOS} conditions, shedding insight into the potential of locating a receiver even when the \ac{LOS} is blocked. We have derived fundamental performance bounds on the estimation uncertainty for delay, angle of arrival, angle of departure, and channel gain of each path, as well as the user position and orientation angle. We also proposed a novel three stage algorithm for the estimation of the user position and orientation angle. This algorithm determines coarse estimates of the channel parameters by exploiting the sparsity of the \ac{mm-wave} in beamspace, followed by an iterative refinement, and finally a conversion to position and orientation. Through simulation studies, we demonstrate the efficiency of the proposed algorithm, and show that even in \ac{OLOS} conditions, it is possible to estimate the user's position and orientation angle, by exploiting the information coming from the multipath, though at a significant performance penalty.

% under  \ac{mm-wave} system. Using LOS and NLOS channel models, we have first computed the \ac{FIM} for delay, AOD, AOA, and channel gain. We have then transformed the FIM to comprise the position and rotation angle of MS. 
%We have proved that obtaining a non-singular FIM with one BS in a LOS in the presence of NLOS \ac{mm-wave} system is possible by using multi-beam transmission and appropriate signal design. 
%Moreover, we have shown that a wideband system can be considered as a system with sequential transmission in frequency that leads to a non-singular FIM with one BS even for $M_{t}=1$ provided that $N>1$ and a special condition on the transmitted signal. 
%A two stage algorithm including the detection phase and the estimation phase for the sparse estimation of \ac{AOA}, \ac{AOD}, and \ac{TOA} has been developed. The results show that \ac{RMSE} of the \ac{AOA}, \ac{AOD}, and \ac{TOA} converge to the corresponding values obtained from the inverse of the \ac{FIM}. Moreover, the estimated values of position and rotation angle converge to the PEB and REB after a few iterations.
\appendices
\section{Elements in \eqref{Parameters8w} }\label{elements}
Replacing $\mathbf{y}[n]$ from (\ref{Receivedb1}) in (\ref{Parameters5b}), using (\ref{Parameters6ww}), and considering $\mathbb{E}_{\mathbf{y}\vert\boldsymbol{\eta}}[\mathbf{n}[n]]=\mathbf{0}$, we obtain
\begin{equation}\label{appa0}
\Psi(x_{r},x_{s})=\frac{2}{N_{0}}\sum_{n=0}^{N-1}\Re\left\{\frac{\partial\boldsymbol{\mu}^{\mathrm{H}}[n]}{\partial x_{r}}\frac{\partial\boldsymbol{\mu}[n]}{\partial x_{s}}\right\}.
\end{equation}
The elements of the \ac{FIM} are obtained based on \eqref{appa0}. The entry associated with the $n$-th subcarrier is denoted as $\Psi_n(x_{r},x_{s})$, and given by  (for $\left\lbrace\tau_{r}, \tau_{s}\right\rbrace$ and $\left\lbrace\boldsymbol{\theta}_{r}, \boldsymbol{\theta}_{s}\right\rbrace$)
%The sub-matrix $\mathbf{\Psi}(\boldsymbol{\eta}_{r},\boldsymbol{\eta}_{s})$ in \eqref{Parameters6w} can be written as
%\begin{equation}\label{Parameters7w}
%\mathbf{\Psi}(\boldsymbol{\eta}_{r},\boldsymbol{\eta}_{s})=\sum_{n=0}^{N-1}
%\mathbf{\Psi}_{n}(\boldsymbol{\eta}_{r},\boldsymbol{\eta}_{s}),
%\end{equation}
%The terms including $\left\lbrace\tau_{r}, \tau_{s}\right\rbrace$ and $\left\lbrace\boldsymbol{\theta}_{r}, \boldsymbol{\theta}_{s}\right\rbrace$ are summarized as:
\begin{IEEEeqnarray}{rCl}
%\begin{equation}\label{FIM5}
 \Psi_{n}(\tau_{r},\tau_{s})& = &\frac{2}{N_{0}}\Re\{\tilde{h}^{*}_{r}\tilde{h}_{s} A_{\mathrm{Rx},n}(\theta_{\mathrm{Rx,r}},\theta_{\mathrm{Rx,s}})A^{(2)}_{\mathrm{Tx},\mathbf{F},n}(\tau_{r},\tau_{s},\theta_{\mathrm{Tx,s}},\theta_{\mathrm{Tx,r}})\},\label{Parameters9w}\\
%\end{equation}
%\begin{equation}\label{FIM6}
 \Psi_{n}(\tau_{r},\theta_{\mathrm{Tx},s})& = &\frac{2}{N_{0}}\Re\{j\tilde{h}^{*}_{r}\tilde{h}_{s}A_{\mathrm{Rx},n}(\theta_{\mathrm{Rx,r}},\theta_{\mathrm{Rx,s}})A^{(1)}_{\mathbf{D}_{\mathrm{Tx},s},\mathbf{F},n}(\tau_{r},\tau_{s},\theta_{\mathrm{Tx,s}},\theta_{\mathrm{Tx,r}})\},\label{Parameters10w}\\
%\end{equation}
%\begin{equation}\label{FIM7}
 \Psi_{n}(\tau_{r},\theta_{\mathrm{Rx},s})& = &\frac{2}{N_{0}}\Re\{j\tilde{h}^{*}_{r}\tilde{h}_{s}A_{\mathbf{D}_{\mathrm{Rx,s}},n}(\theta_{\mathrm{Rx,r}},\theta_{\mathrm{Rx,s}})A^{(1)}_{\mathrm{Tx},\mathbf{F},n}(\tau_{r},\tau_{s},\theta_{\mathrm{Tx,s}},\theta_{\mathrm{Tx,r}})\},\label{Parameters11w}\\
%\end{equation}
%\begin{equation}\label{FIM10}
 \Psi_{n}(\theta_{\mathrm{Tx},r},\theta_{\mathrm{Tx},s})& = &\frac{2}{N_{0}}\Re\{\tilde{h}^{*}_{r}\tilde{h}_{s}A_{\mathrm{Rx},n}(\theta_{\mathrm{Rx,r}},\theta_{\mathrm{Rx,s}})A_{\mathbf{Dd}_{\mathrm{Tx}},\mathbf{F},n}(\tau_{r},\tau_{s},\theta_{\mathrm{Tx,s}},\theta_{\mathrm{Tx,r}})\}, \label{Parameters12w}\\
%\end{equation}
%\begin{equation}\label{FIM11}
 \Psi_{n}(\theta_{\mathrm{Tx},r},\theta_{\mathrm{Rx},s})& = &\frac{2}{N_{0}}\Re\{\tilde{h}^{*}_{r}\tilde{h}_{s}A_{\mathbf{D}_{\mathrm{Rx,s}},n}(\theta_{\mathrm{Rx,r}},\theta_{\mathrm{Rx,s}})A^{(0)}_{\mathbf{D}_{\mathrm{Tx},r},\mathbf{F},n}(\tau_{r},\tau_{s},\theta_{\mathrm{Tx,s}},\theta_{\mathrm{Tx,r}})\}, \label{Parameters13w}\\
%\end{equation}
%\begin{equation}\label{FIM14}
  \Psi_{n}(\theta_{\mathrm{Rx},r},\theta_{\mathrm{Rx},s})& = &\frac{2}{N_{0}}\Re\{\tilde{h}^{*}_{r}\tilde{h}_{s}A_{\mathbf{D}_{\mathrm{Rx,r,s}},n}(\theta_{\mathrm{Rx,r}},\theta_{\mathrm{Rx,s}})A^{(0)}_{\mathrm{Tx},\mathbf{F},n}(\tau_{r},\tau_{s},\theta_{\mathrm{Tx,s}},\theta_{\mathrm{Tx,r}})\}.\label{Parameters14w}
%\end{equation}
\end{IEEEeqnarray}
The following notations are introduced:
\begin{align}
A^{(k)}_{\mathrm{Tx},\mathbf{F},n}(\tau_{r},\tau_{s},\theta_{\mathrm{Tx,s}},\theta_{\mathrm{Tx,r}})&\triangleq\mathbf{a}_{\mathrm{Tx},\mathbf{F},n}^{\mathrm{H}}(\theta_{\mathrm{Tx,s}})\mathbf{A}_{k,n}(\tau_{r},\tau_{s})\mathbf{a}_{\mathrm{Tx},\mathbf{F},n}(\theta_{\mathrm{Tx,r}}),\label{Parameters15w}\\
A^{(l)}_{\mathbf{D}_{\mathrm{Tx},s},\mathbf{F},n}(\tau_{r},\tau_{s},\theta_{\mathrm{Tx,s}},\theta_{\mathrm{Tx,r}})&\triangleq\mathbf{a}_{\mathbf{D}_{\mathrm{Tx}},\mathbf{F},n}^{\mathrm{H}}(\theta_{\mathrm{Tx,s}})\mathbf{A}_{l,n}(\tau_{r},\tau_{s})\mathbf{a}_{\mathrm{Tx},\mathbf{F},n}(\theta_{\mathrm{Tx,r}}),\label{Parameters16w}\\
A^{(l)}_{\mathbf{D}_{\mathrm{Tx},r},\mathbf{F},n}(\tau_{r},\tau_{s},\theta_{\mathrm{Tx,s}},\theta_{\mathrm{Tx,r}})&\triangleq\mathbf{a}_{\mathrm{Tx},\mathbf{F},n}^{\mathrm{H}}(\theta_{\mathrm{Tx,s}})\mathbf{A}_{l,n}(\tau_{r},\tau_{s})\mathbf{a}_{\mathbf{D}_{\mathrm{Tx}},\mathbf{F},n}(\theta_{\mathrm{Tx,r}}),\label{Parameters17w}\\
A_{\mathbf{Dd}_{\mathrm{Tx}},\mathbf{F},n}(\tau_{r},\tau_{s},\theta_{\mathrm{Tx,s}},\theta_{\mathrm{Tx,r}})&\triangleq\mathbf{a}_{\mathbf{D}_{\mathrm{Tx}},\mathbf{F},n}^{\mathrm{H}}(\theta_{\mathrm{Tx,s}})\mathbf{A}_{0,n}(\tau_{r},\tau_{s})\mathbf{a}_{\mathbf{D}_{\mathrm{Tx}},\mathbf{F},n}(\theta_{\mathrm{Tx,r}}),\label{Parameters18w}
\end{align}
where $l \in \{0,1\}$, and $\mathbf{A}_{k,n}(\tau_{r},\tau_{s}), k \in \{0,1,2 \}$, is given by
\vspace{-2mm}
\begin{equation}\label{Parameters19w}
\mathbf{A}_{k,n}(\tau_{r},\tau_{s})\triangleq (2\pi n/(NT_{s}))^{k}~\mathbf{x}[n]\mathbf{x}^{\mathrm{H}}[n]e^{-j2\pi n(\tau_{r}-\tau_{s})/(NT_{s})}.
\end{equation}
The vectors $\mathbf{a}_{\mathrm{Tx},\mathbf{F},n}(\theta_{\mathrm{Tx},r})$ and $\mathbf{a}_{\mathbf{D}_{\mathrm{Tx}},\mathbf{F},n}(\theta_{\mathrm{Tx},r})$ are given by $\mathbf{a}_{\mathrm{Tx},\mathbf{F},n}(\theta_{\mathrm{Tx},r}) = \mathbf{F}^{\mathrm{H}}[n]\mathbf{a}_{\mathrm{Tx},n}(\theta_{\mathrm{Tx},r})$ and $\mathbf{a}_{\mathbf{D}_{\mathrm{Tx}},\mathbf{F},n}(\theta_{\mathrm{Tx},r})=\mathbf{F}^{\mathrm{H}}[n]\mathbf{D}_{\mathrm{Tx},r}[n]\mathbf{a}_{\mathrm{Tx},n}(\theta_{\mathrm{Tx},r})$. The matrix $\mathbf{D}_{\mathrm{Tx},r}[n]$ is defined as
\begin{equation}\label{Parameters19ww}
\mathbf{D}_{\mathrm{Tx},r}[n]\triangleq j\frac{2\pi}{\lambda_{n}}d\cos(\theta_{\mathrm{Tx},r})\mathrm{diag}\{0,\ldots,N_{\mathrm{t}}-1\}.
\end{equation}
The scalars $A_{\mathrm{Rx},n}(\theta_{\mathrm{Rx,r}},\theta_{\mathrm{Rx,s}})$, $A_{\mathbf{D}_{\mathrm{Rx,s}},n}(\theta_{\mathrm{Rx,r}},\theta_{\mathrm{Rx,s}})$, and $A_{\mathbf{D}_{\mathrm{Rx,r,s}},n}(\theta_{\mathrm{Rx,r}},\theta_{\mathrm{Rx,s}})$ are defined as
\begin{align}
 A_{\mathrm{Rx},n}(\theta_{\mathrm{Rx,r}},\theta_{\mathrm{Rx,s}})&\triangleq\mathbf{a}^{\mathrm{H}}_{\mathrm{Rx},n}(\theta_{\mathrm{Rx},r})\mathbf{a}_{\mathrm{Rx},n}(\theta_{\mathrm{Rx},s}),
\label{Parameters20w}\\
 A_{\mathbf{D}_{\mathrm{Rx,s}},n}(\theta_{\mathrm{Rx,r}},\theta_{\mathrm{Rx,s}})&\triangleq\mathbf{a}^{\mathrm{H}}_{\mathrm{Rx},n}(\theta_{\mathrm{Rx},r})\mathbf{D}_{\mathrm{Rx},s}[n]\mathbf{a}_{\mathrm{Rx},n}(\theta_{\mathrm{Rx},s}),
\label{Parameters21w}\\
 A_{\mathbf{D}_{\mathrm{Rx,r,s}},n}(\theta_{\mathrm{Rx,r}},\theta_{\mathrm{Rx,s}})&\triangleq\mathbf{a}^{\mathrm{H}}_{\mathrm{Rx},n}(\theta_{\mathrm{Rx},r})\mathbf{D}^{\mathrm{H}}_{\mathrm{Rx},r}[n]\mathbf{D}_{\mathrm{Rx},s}[n]\mathbf{a}_{\mathrm{Rx},n}(\theta_{\mathrm{Rx},s}),
\label{Parameters22w}
\end{align}
where $\mathbf{D}_{\mathrm{Rx},r}[n]$ has the same expression as \eqref{Parameters19ww} by replacing the subscript $\mathrm{Tx}$ by $\mathrm{Rx}$ and $N_{t}$ by $N_{r}$. The terms including channel coefficients are summarized as:
\begin{multline}\label{appa1}
 \mathbf{\Psi}_{n}(\tau_{r},\tilde{\mathbf{h}}_{s})= \frac{2}{N_{0}}[\Re\{j\tilde{h}^{*}_{r}A_{\mathrm{Rx},n}(\theta_{\mathrm{Rx,r}},\theta_{\mathrm{Rx,s}})A^{(1)}_{\mathrm{Tx},\mathbf{F},n}(\tau_{r},\tau_{s},\theta_{\mathrm{Tx},s},\theta_{\mathrm{Tx},r})\},\\\Re\{-\tilde{h}^{*}_{r}A_{\mathrm{Rx},n}(\theta_{\mathrm{Rx,r}},\theta_{\mathrm{Rx,s}})A^{(1)}_{\mathrm{Tx},\mathbf{F},n}(\tau_{r},\tau_{s},\theta_{\mathrm{Tx},s},\theta_{\mathrm{Tx},r})\}],
\end{multline}
\begin{multline}\label{appa2}
 \mathbf{\Psi}_{n}(\theta_{\mathrm{Tx},r},\tilde{\mathbf{h}}_{s})= \frac{2}{N_{0}}[\Re\{\tilde{h}^{*}_{r}A_{\mathrm{Rx},n}(\theta_{\mathrm{Rx,r}},\theta_{\mathrm{Rx,s}})A^{(0)}_{\mathbf{D}_{\mathrm{Tx},r},\mathbf{F},n}(\tau_{r},\tau_{s},\theta_{\mathrm{Tx,s}},\theta_{\mathrm{Tx,r}})\},\\\Re\{j\tilde{h}^{*}_{r}A_{\mathrm{Rx},n}(\theta_{\mathrm{Rx,r}},\theta_{\mathrm{Rx,s}})A^{(0)}_{\mathbf{D}_{\mathrm{Tx},r},\mathbf{F},n}(\tau_{r},\tau_{s},\theta_{\mathrm{Tx,s}},\theta_{\mathrm{Tx,r}})\}],
\end{multline}
\begin{multline}\label{appa3}
 \mathbf{\Psi}_{n}(\theta_{\mathrm{Rx},r},\tilde{\mathbf{h}}_{s})= -\frac{2}{N_{0}}[\Re\{\tilde{h}^{*}_{r}A_{\mathbf{D}_{\mathrm{Rx},r},n}(\theta_{\mathrm{Rx,r}},\theta_{\mathrm{Rx,s}})A^{(0)}_{\mathrm{Tx},\mathbf{F},n}(\tau_{r},\tau_{s},\theta_{\mathrm{Tx},s},\theta_{\mathrm{Tx},r})\},\\\Re\{j\tilde{h}^{*}_{r}A_{\mathbf{D}_{\mathrm{Rx},r},n}(\theta_{\mathrm{Rx,r}},\theta_{\mathrm{Rx,s}})A^{(0)}_{\mathrm{Tx},\mathbf{F},n}(\tau_{r},\tau_{s},\theta_{\mathrm{Tx},s},\theta_{\mathrm{Tx},r})\}],
\end{multline}
\begin{multline}\label{appa4}
 \Psi_{n}(\Re\{\tilde{h}_{r}\},\Re\{\tilde{h}_{s}\})=\Psi_{n}(\Im\{\tilde{h}_{r}\},\Im\{\tilde{h}_{s}\})=\\ \frac{2}{N_{0}}\Re\{A_{\mathrm{Rx},n}(\theta_{\mathrm{Rx,r}},\theta_{\mathrm{Rx,s}})A^{(0)}_{\mathrm{Tx},\mathbf{F},n}(\tau_{r},\tau_{s},\theta_{\mathrm{Tx},s},\theta_{\mathrm{Tx},r})\},
\end{multline}
\begin{multline}\label{appa5}
 \Psi_{n}(\Re\{\tilde{h}_{r}\},\Im\{\tilde{h}_{s}\})=-\Psi_{n}(\Im\{\tilde{h}_{r}\},\Re\{\tilde{h}_{s}\})=\\ \frac{2}{N_{0}}\Re\{jA_{\mathrm{Rx},n}(\theta_{\mathrm{Rx,r}},\theta_{\mathrm{Rx,s}})A^{(0)}_{\mathrm{Tx},\mathbf{F},n}(\tau_{r},\tau_{s},\theta_{\mathrm{Tx},s},\theta_{\mathrm{Tx},r})\}.
\end{multline}
\section{Complexity Analysis}\label{comprep}
We analyze the complexity of different stages of the proposed algorithm.\begin{itemize}\item Coarse Estimation: The complexity in performing \eqref{tinex1ee} is on the order of $O(N^{2}_{r}N^{2}_{t}GN_{\mathrm{sub}})$ where $N_{\mathrm{sub}}$ denotes the few subcarriers sufficient to detect the dominant path. The QR factorization of the mutilated basis $\mathbf{\Omega}_{\mathcal{K}_{t}}[n]$ approximately requires $O(GN_{r}\hat{K}^{2})$ operations for each subcarrier, and matrix inversion to obtain the channel coefficients in \eqref{BWTransceiver2z} approximately takes $O(N\hat{K}^{3})$ operations for all the subcarriers. The complexity in computing \eqref{BWTransceiver2tzsfzxe3} is on the order of $O(ND_{o}\hat{K})$ where $D_{o}$ denotes the number of delay grid points, and \eqref{BWTransceiver2tzsfzxe2} requires $O(N\hat{K})$ operations. Consequently, the maximum complexity from coarse estimation of the channel parameters is dominated by the term $\hat{K}\times O(N^{2}_{r}N^{2}_{t}GN_{\mathrm{sub}})$.
%The complexity in performing \eqref{tinex1ee} is of the order of $O(N^{2}_{r}N^{2}_{t}G)$ for each subcarriers. Since it is sufficient in principle to detect the dominant path with a few subcarriers, the complexity in computing \eqref{tinex1ee} would be around $O(N^{2}_{r}N^{2}_{t}GN_{\mathrm{sub}})$ where $N_{\mathrm{sub}}$ denotes the few subcarriers sufficient to detect the dominant paths. 
%The QR factorization and matrix inversion in the later steps of the modified DCS-SOMP algorithm to obtain the channel coefficients for all the subcarriers in \eqref{BWTransceiver2z} approximately takes $O(N_{r}N_{t}G\hat{K}^{2}N)$ operations. After detection of different paths using the modified DCS-SOMP algorithm, the complexity in computing \eqref{BWTransceiver2tzsfzxe3} is on the order of $O(ND_{o}\hat{K})$ where $D_{o}$ denotes the number of delay grid points, and \eqref{BWTransceiver2tzsfzxe2} requires $O(N\hat{K})$ operations. Consequently, the maximum complexity from coarse estimation of the channel parameters is dominated by the term $\hat{K}\times O(N^{2}_{r}N^{2}_{t}GN_{\mathrm{sub}})$.
\item Fine Estimation: In the refinement phase, the complexity is mainly affected by Gauss-Seidel-type iterations with first and second order derivatives of a vector $\mathbf{a}(x)$ of length $L_{\mathrm{x}}$ with respect to a variable $x$ that can be delay, AOA, and AOD. These operations lead to a complexity on the order of $O(L^{2}_{\mathrm{x}}N)$ for each path. Given the subsequent path refinement, the maximum complexity of fine estimation is on the order of $O(\hat{K}^{2})\times O(L^{2}_{\mathrm{x}}N)$.\item Conversion to Position and Orientation: The conversion to position and orientation in the LOS case is easy to implement since it involves only some basic operations. For the NLOS and OLOS scenarios, the LMA algorithm is applied. It is not considered the complexity driver, since it combines the advantages of gradient-descent and Gauss-Newton methods. The LMA algorithm can be effectively applied by implementing delayed gratification, which leads to higher success rate and fewer Jacobian evaluations.
\end{itemize}
%Given the explanations about the complexity of different stages of the proposed algorithm in Appendix C. It is true that the proposed approach is complex; however, it applies well implemented algorithms such as DCS-SOMP that can be efficiently used in practice. Moreover, one can use the proposed estimation method in the uplink that allows the BS to solve a complex algorithm for user localization. And in any case, the objective of the paper is to present the bounds and some algorithms that approach the bounds, focusing on performance not on complexity. Whether simpler algorithms that also approach the bounds can be derived is an interesting topic for further research.
\bibliographystyle{IEEEtran}
\bibliography{Extended_Refs.bib}

\begin{thebibliography}{10}
\providecommand{\url}[1]{#1}
\csname url@samestyle\endcsname
\providecommand{\newblock}{\relax}
\providecommand{\bibinfo}[2]{#2}
\providecommand{\BIBentrySTDinterwordspacing}{\spaceskip=0pt\relax}
\providecommand{\BIBentryALTinterwordstretchfactor}{4}
\providecommand{\BIBentryALTinterwordspacing}{\spaceskip=\fontdimen2\font plus
\BIBentryALTinterwordstretchfactor\fontdimen3\font minus
  \fontdimen4\font\relax}
\providecommand{\BIBforeignlanguage}[2]{{%
\expandafter\ifx\csname l@#1\endcsname\relax
\typeout{** WARNING: IEEEtran.bst: No hyphenation pattern has been}%
\typeout{** loaded for the language `#1'. Using the pattern for}%
\typeout{** the default language instead.}%
\else
\language=\csname l@#1\endcsname
\fi
#2}}
\providecommand{\BIBdecl}{\relax}
\BIBdecl

\bibitem{Arash}
A.~Shahmansoori, G.~Garcia, G.~Destino, G.~Seco-Granados, and H.~Wymeersch,
  ``5{G} position and orientation estimation through millimeter wave {MIMO},''
  in \emph{Proc. IEEE Globecom}, Dec 2015.

\bibitem{Zhouyue}
P.~Zhouyue and F.~Khan, ``An introduction to millimeter-wave mobile broadband
  systems,'' \emph{IEEE Communications Magazine}, vol.~49, no.~6, pp. 101--107,
  2011.

\bibitem{Rappaport}
T.~Rappaport, S.~Sun, R.~Mayzus, H.~Zhao, Y.~Azar, K.~Wang, G.~Wong, J.~Schulz,
  M.~Samimi, and F.~Gutierrez, ``Millimeter wave mobile communications for {5G}
  cellular: It will work!'' \emph{IEEE Access}, vol.~1, pp. 335--349, 2013.

\bibitem{Wang}
J.~Wang, ``Beam codebook based beamforming protocol for multi-{Gbps}
  millimeter-wave {WPAN} systems,'' \emph{IEEE J. Sel. Areas Commun.}, vol.~27,
  no.~8, pp. 1390--1399, 2009.

\bibitem{Hur}
S.~Hur, T.~Kim, D.~Love, J.~Krogmeier, T.~Thomas, and A.~Ghosh, ``Millimeter
  wave beamforming for wireless backhaul and accessin small cell networks,''
  \emph{IEEE Trans. Commun.}, vol.~61, no.~10, pp. 4391--4403, 2013.

\bibitem{Tsang}
Y.~Tsang, A.~Poon, and S.~Addepalli, ``Coding the beams: Improving beamforming
  training in mmwave communication system,'' in \emph{Global Telecomm. Conf.
  (GLOBECOM)}, 2011.

\bibitem{Duarte}
M.~F. Duarte, S.~Sarvotham, D.~Baron, M.~B. Wakin, and R.~G. Baraniuk,
  ``Distributed compressed sensing of jointly sparse signals,'' in \emph{in
  Proc. 39th Asilomar Conf. Sig., Syst., Comp.}, 2005, pp. 3469--3472.

\bibitem{Duarte2}
M.~F. Duarte, V.~Cevher, and R.~G. Baraniuk, ``Model-based compressive sensing
  for signal ensembles,'' in \emph{in Proc. 47th Ann. Allerton Conf.
  Communication, Control, Computing, (Monticello, IL)}, 2009, pp. 244--250.

\bibitem{Bolcskei}
Y.~C. Eldar, P.~Kuppinger, and Bolcskei, ``Block-sparse signals: Uncertainty
  relations and efficient recovery,'' \emph{IEEE Trans. Signal Processing},
  vol.~58, no.~6, pp. 3042--3054, 2010.

\bibitem{BspaceSayeed}
J.~Brady, N.~Behdad, and A.~Sayeed, ``Beamspace {MIMO} for millimeter-wave
  communications: System architecture, modeling, analysis, and measurements,''
  \emph{IEEE Transactions on Antennas and Propagation}, vol.~61, no.~7, pp.
  3814--3827, 2013.

\bibitem{widebandbrady}
J.~H. Brady and A.~Sayeed, ``Wideband communication with high-dimensional
  arrays: New results and transceiver architectures,'' in \emph{IEEE
  International Conference on Communication (ICC)}, 2015.

\bibitem{Martinez-Ingles}
M.~T. Martinez-Ingles, D.~P. Gaillot, J.~Pascual-Garcia, J.~M.
  Molina-Garcia-Pardo, M.~Lienard, and J.~V. Rodriguez, ``Deterministic and
  experimental indoor {mmW} channel modeling,'' \emph{IEEE Ant. Wireless Prop.
  Lett.}, vol.~13, pp. 1047--1050, May. 2014.

\bibitem{Vaughan}
R.~G. Vaughan and J.~B. Andersen, \emph{Channels, Propagation and Antennas for
  Mobile Communications}.\hskip 1em plus 0.5em minus 0.4em\relax London, UK:
  Institute of Electrical Engineers (IEE), 2003.

\bibitem{mmMAGIC}
mmMAGIC White Paper~2.1, ``Measurement campaigns and initial channel models for
  preferred suitable frequency ranges,'' \emph{https://5g-mmmagic.eu/}, vol.
  version 1.0, Mar. 2016.

\bibitem{Marzi}
Z.~Marzi, D.~Ramasamy, and U.~Madhow, ``Compressive channel estimation and
  tracking for large arrays in mm-wave picocells,'' \emph{IEEE Journal of
  Selected Topics in Signal Processing}, vol.~10, no.~3, pp. 514--527, Apr.
  2016.

\bibitem{LeeJ}
J.~Lee, G.-T. Gil, and Y.~H. Lee, ``Channel estimation via orthogonal matching
  pursuit for hybrid {MIMO} systems in millimeter wave communications,''
  \emph{IEEE Transactions on Communications}, vol.~64, no.~6, pp. 2370--2386,
  Jun. 2016.

\bibitem{AlkhateebA}
A.~Alkhateeb, O.~E. Ayach, G.~Leus, and R.~W. Heath~Jr., ``Channel estimation
  and hybrid precoding for millimeter wave cellular systems,'' \emph{IEEE
  Journal of Selected Topics in Signal Processing}, vol.~8, no.~5, pp.
  831--846, Oct. 2014.

\bibitem{ChoiJ}
J.~Choi, ``Beam selection in mm-wave multiuser {MIMO} systems using compressive
  sensing,'' \emph{IEEE Transactions on Communications}, vol.~63, no.~8, pp.
  2936--2947, Aug. 2015.

\bibitem{AlkhateebC}
A.~Alkhateeb, O.~E. Ayach, G.~Leus, and R.~W. Heath~Jr., ``Compressed-sensing
  based multi-user millimeter wave systems: How many measurements are needed?''
  in \emph{Proc. IEEE Int. Conf. Acoustics, Speech and Sig. Process.
  (ICASSP)}.\hskip 1em plus 0.5em minus 0.4em\relax Brisbane, Australia: arXiv
  preprint arXiv:1505.00299, Apr 2015.

\bibitem{HanY}
Y.~Han and J.~Lee, ``Two-stage compressed sensing for millimeter wave channel
  estimation,'' in \emph{Proc. IEEE Int. Symp. on Inform. Theory (ISIT)}, 2016,
  pp. 860--864.

\bibitem{LeeJ2}
J.~Lee, G.-T. Gil, and Y.~Lee, ``Exploiting spatial sparsity for estimating
  channels of hybrid {MIMO} systems in millimeter wave communications,'' in
  \emph{Proc. IEEE Global Telecommun. Conf. (GLOBECOM)}, Dec 2014, pp.
  3326--3331.

\bibitem{Ramasamy}
D.~Ramasamy, S.~Venkateswaran, and U.~Madhow, ``Compressive adaptation of large
  steerable arrays,'' in \emph{Proc. IEEE Inform. Theory and Applicat. Workshop
  (ITA)}, Feb 2012, pp. 234--239.

\bibitem{BerrakiD}
D.~E. Berraki, S.~M.~D. Armour, and A.~R. Nix, ``Application of compressive
  sensing in sparse spatial channel recovery for beamforming in mmwave outdoor
  systems,'' in \emph{Proc. IEEE Wireless Commun. and Networking Conf.}, Apr
  2014, pp. 887--892.

\bibitem{sanchis2002novel}
P.~Sanchis, J.~Martinez, J.~Herrera, V.~Polo, J.~Corral, and J.~Marti, ``A
  novel simultaneous tracking and direction of arrival estimation algorithm for
  beam-switched base station antennas in millimeter-wave wireless broadband
  access networks,'' in \emph{IEEE Antennas and Propagation Society
  International Symposium}, 2002.

\bibitem{DenSaya}
H.~Deng and A.~Sayeed, ``Mm-wave {MIMO} channel modeling and user localization
  using sparse beamspace signatures,'' in \emph{International Workshop on
  Signal Processing Advances in Wireless Communications}, 2014, pp. 130--134.

\bibitem{vari2014mmwaves}
M.~Vari and D.~Cassioli, ``{mmWaves RSSI indoor network localization},'' in
  \emph{ICC Workshop on Advances in Network Localization and Navigation}, 2014.

\bibitem{hu2014esprit}
A.~Hu, T.~Lv, H.~Gao, Z.~Zhang, and S.~Yang, ``An {ESPRIT}-based approach for
  {2-D} localization of incoherently distributed sources in massive {MIMO}
  systems,'' \emph{IEEE Journal of Selected Topics in Signal Processing},
  vol.~8, no.~5, pp. 996--1011, 2014.

\bibitem{Dardari}
A.~Guerra, F.~Guidi, and D.~Dardari, ``Position and orientation error bound for
  wideband massive antenna arrays,'' in \emph{ICC Workshop on Advances in
  Network Localization and Navigation}, 2015.

\bibitem{savic2015fingerprinting}
V.~Savic and E.~G. Larsson, ``Fingerprinting-based positioning in distributed
  massive {MIMO} systems,'' in \emph{IEEE Vehicular Technology Conference},
  2015.

\bibitem{NGarcia}
N.~Garcia, H.~Wymeersch, E.~G. Str{\"o}m, and D.~Slock, ``Location-aided
  mm-wave channel estimation for vehicular communication,'' in \emph{Proc. IEEE
  International Workshop on Signal Processing Advances in Wireless
  Communications (SPAWC)}, Edinburgh, England, 3-6 Jul 2016.

\bibitem{NGarcia2}
N.~Garcia, H.~Wymeersch, E.~G. Larsson, A.~M. Haimovich, and M.~Coulon,
  ``Direct localization for massive {MIMO},'' \emph{IEEE Transactions on Signal
  Processing}, vol.~65, no.~10, pp. 2475--2487, Feb. 2017.

\bibitem{linhyb}
J.~Li, J.~Conan, and S.~Pierre, ``Position location of mobile terminal in
  wireless {MIMO} communication systems,'' \emph{Journal of Communications and
  Networks}, vol.~9, no.~3, pp. 254--264, Sep. 2007.

\bibitem{DBLP:journals/twc/KoivistoCWHTLKV17}
\BIBentryALTinterwordspacing
M.~Koivisto, M.~Costa, J.~Werner, K.~Heiska, J.~Talvitie, K.~Lepp{\"{a}}nen,
  V.~Koivunen, and M.~Valkama, ``Joint device positioning and clock
  synchronization in 5{G} ultra-dense networks,'' \emph{{IEEE} Trans. Wireless
  Communications}, vol.~16, no.~5, pp. 2866--2881, 2017. [Online]. Available:
  \url{https://doi.org/10.1109/TWC.2017.2669963}
\BIBentrySTDinterwordspacing

\bibitem{DBLP:journals/corr/KoivistoHCKLV16}
\BIBentryALTinterwordspacing
M.~Koivisto, A.~Hakkarainen, M.~Costa, P.~Kela, K.~Lepp{\"{a}}nen, and
  M.~Valkama, ``High-efficiency device positioning and location-aware
  communications in dense 5{G} networks,'' \emph{{IEEE} Communications
  Magazine}, vol.~55, no.~8, pp. 188--195, 2017. [Online]. Available:
  \url{http://ieeexplore.ieee.org/document/7984759/}
\BIBentrySTDinterwordspacing

\bibitem{Stoicapp}
P.~Stoica and T.~S\"{o}derstr\"{o}m, ``On reparametrization of loss functions
  used in estimation and the invariance principle,'' \emph{Signal Process.},
  vol.~17, pp. 383--387, 1989.

\bibitem{Swindlehurstt}
A.~L. Swindlehurst and P.~Stoica, ``Maximum likelihood methods in radar array
  signal processing,'' \emph{Proceedings of the IEEE}, vol.~86, no.~2, pp.
  421--441, 2002.

\bibitem{khateeb3}
A.~Alkhateeb and R.~W. Heath, ``Frequency selective hybrid precoding for
  limited feedback millimeter wave systems,'' \emph{IEEE Transactions on
  Communications}, vol.~64, no.~5, pp. 1801--1818, Oct. 2016.

\bibitem{Kay}
S.~M. Kay, \emph{Fundamentals of Statistical Signal Processing: Estimation
  Theory}.\hskip 1em plus 0.5em minus 0.4em\relax New York, NY, USA: Prentice
  Hall, 2010.

\bibitem{Poor}
H.~V. Poor, \emph{An Introduction to Signal Detection and Estimation, 2nd
  ed.}\hskip 1em plus 0.5em minus 0.4em\relax New York: Springer-Verlag, 1994.

\bibitem{LeitingerJSAC2015}
E.~Leitinger, P.~Meissner, C.~Rudisser, G.~Dumphart, and K.~Witrisal,
  ``Evaluation of position-related information in multipath components for
  indoor positioning,'' \emph{IEEE J. Sel. Areas Commun.}, vol.~33, no.~11,
  Nov. 2015.

\bibitem{WitrisalSPM2016}
K.~Witrisal, P.~Meissner, E.~Leitinger, Y.~Shen, C.~Gustafson, F.~Tufvesson,
  K.~Haneda, D.~Dardari, A.~F. Molisch, A.~Conti, and M.~Z. Win,
  ``High-accuracy localization for assisted living: {5G} systems will turn
  multipath channels from foe to friend,'' \emph{IEEE Signal Process. Mag.},
  vol.~33, no.~2, Mar. 2016.

\bibitem{DBLP:journals/corr/Abu-ShabanZASW17}
\BIBentryALTinterwordspacing
Z.~Abu{-}Shaban, X.~Zhou, T.~D. Abhayapala, G.~Seco{-}Granados, and
  H.~Wymeersch, ``Error bounds for uplink and downlink 3{D} localization in
  5{G} mmwave systems,'' \emph{CoRR}, vol. abs/1704.03234, 2017. [Online].
  Available: \url{http://arxiv.org/abs/1704.03234}
\BIBentrySTDinterwordspacing

\bibitem{LeitingerPhD2016}
E.~Leitinger, ``Cognitive indoor positioning and tracking using multipath
  channel information,'' Ph.D. dissertation, Graz University of Technology,
  2016.

\bibitem{BspaceSayeedx}
A.~Sayeed, ``Deconstructing multiantenna fading channels,'' \emph{IEEE
  Transactions on Signal Processing}, vol.~50, no.~10, pp. 2563--2579, 2002.

\bibitem{Davies}
M.~E. Davies and Y.~C. Eldar, ``Rank awareness in joint sparse recovery,''
  \emph{IEEE Transactions on Information Theory}, vol.~58, no.~2, pp.
  1135--1146, Feb. 2012.

\bibitem{Ortega}
J.~Ortega and M.~Rockoff, ``Nonlinear difference equations and gauss-seidel
  type iterative methods,'' \emph{SIAM J. Numer. Anal.}, vol.~3, no.~3, pp.
  497--513, 1966.

\bibitem{Zacks}
S.~Zacks, \emph{Parametric Statistical Inference: Basic Theory and Modern
  Approaches}.\hskip 1em plus 0.5em minus 0.4em\relax Oxford, England:
  Pergamon, 1981.

\bibitem{Levenberg}
K.~Levenberg, ``A method for the solution of certain non-linear problems in
  least squares,'' \emph{Quarterly of Applied Mathematics}, vol.~2, pp.
  164--168, 1944.

\bibitem{Marquardt}
D.~Marquardt, ``An algorithm for least-squares estimation of nonlinear
  parameters,'' \emph{SIAM Journal on Applied Mathematics}, vol.~11, no.~2, pp.
  431--441, 1963.

\bibitem{Maltsevx}
K.~e.~a. Maltsev, ``{IEEE} doc. 802.11-08/1044r0. 60 {GH}z {WLAN} experimental
  investigations,'' Sep. 2008.

\bibitem{geomted}
Q.~C. Li, G.~Wu, and T.~S. Rappaport, ``Channel model for millimeterwave
  communications based on geometry statistics,'' in \emph{IEEE Globecom
  Workshop}, 2014.

\bibitem{Qian1}
Q.~C. Li, H.~Shirani-Mehr, T.~Balercia, A.~Papathanassiou, G.~Wu, S.~Sun, M.~K.
  Samimi, and T.~S. Rappaport, ``Validation of a geometry-based statistical
  mmwave channel model using ray-tracing simulation,'' in \emph{IEEE Vehicular
  Technology Conference}, May. 2015, pp. 1--5.

\bibitem{Qian2}
Q.~C. Li, G.~Wu, and T.~S. Rappaport, ``Channel model for millimeter wave
  communications based on geometry statistics,'' in \emph{IEEE Globecom
  Workshop}, Dec. 2014, pp. 427--432.

\bibitem{newref5gsix}
A.~Maltsev, R.~Maslennikov, A.~Sevastyanov, A.~Khoryaev, and A.~Lomayev,
  ``Experimental investigations of 60 {GHz} {WLAN} systems in office
  environment,'' \emph{IEEE Journal on Selected Areas in Communications},
  vol.~27, no.~8, pp. 1488--1499, Oct. 2009.

\end{thebibliography}
\end{document}